\documentclass[onecolumn]{aastex631}

\usepackage{breqn}

\shorttitle{Distant RR Lyrae Kinematics}
\shortauthors{Feng et al.}
\graphicspath{{./}{figures/}}
\begin{document}

%\title{The Next Generation Virgo Cluster Survey. XXXVIII.\\ Kinematics of Distant Milky Way RR Lyrae Stars out to 150~kpc}

\title{Kinematics of Distant Milky Way Halo RR Lyrae Stars out to 160~kpc}

\correspondingauthor{Yuting Feng}
\email{yfeng47@ucsc.edu}

% change this to Yuting's ORCID
\author{Yuting Feng}
\affiliation{Department of Astronomy and Astrophysics, University of California Santa Cruz, 1156 High Street, Santa Cruz, CA 95064, USA}

\author[0000-0001-8867-4234]{Puragra Guhathakurta}
\affiliation{Department of Astronomy and Astrophysics, University of California Santa Cruz, 1156 High Street, Santa Cruz, CA 95064, USA}

\author[0000-0002-2073-2781]{Eric W.\ Peng}
\affiliation{National Optical-Infrared Astronomy Research Laboratory (NOIRLab), 950 North Cherry Avenue, Tucson, AZ 85719, USA}

\author[0000-0002-6993-0826]{Emily C.\ Cunningham}
\affiliation{Department of Astronomy, Boston University, 685 Commonwealth Ave, Boston, MA 02215, USA}

\author[0000-0003-1184-8114]{Patrick Côté}
\affiliation{NRC Herzberg Astronomy and Astrophysics, 5071 West Saanich Road, Victoria, BC, V9E 2E7, Canada}

\author[0000-0002-8224-1128]{Laura Ferrarese}
\affiliation{NRC Herzberg Astronomy and Astrophysics, 5071 West Saanich Road, Victoria, BC, V9E 2E7, Canada}

\author[0000-0001-8221-8406]{Stephen D.\ J.\ Gwyn}
\affiliation{NRC Herzberg Astronomy and Astrophysics, 5071 West Saanich Road, Victoria, BC, V9E 2E7, Canada}



\begin{abstract}

We present a kinematical study of the outer halo ($r_{\rm GC}\sim60$--160~kpc) of the Milky Way (MW) based on spectroscopy of 55 RR Lyrae stars using the ESI instrument on the Keck~II telescope. Our spectroscopic targets were selected from three photometric surveys: NGVS, DES, and Pan-STARRS~1. We derive center-of-mass radial velocities with uncertainties of 6--35~km~s$^{-1}$. The halo velocity dispersion measured with our sample is $70 \pm 7$~km~s$^{-1}$. The velocity field shows a possible dipole-like structure, with redshifted northern and blueshifted southern hemispheres. Fitting a MW–Large Magellanic Cloud (LMC) dipole perturbation model yields a weak/marginal dipole signal: amplitude of $-30^{+16}_{-20}$~km~s$^{-1}$ and apex direction $(l,b)=(-38.2^{+42.4}_{-31.5}, -41.3^{+27.9}_{-23.8})$~deg, along with a bulk compression velocity of $-16 \pm 11$~km~s$^{-1}$. While limited by sky coverage and sample size, our results are consistent with the presence of LMC-induced disequilibrium in the distant halo beyond 100 kpc. Aside from the 55 RR~Lyrae stars, our spectroscopic analysis reveals that 10 additional phometrically-selected RR Lyrae candidates are, in fact, quasar/blazar contaminants; this provides a cautionary tale about the presence of such contaminants in sparsely-sampled photometric surveys. Our study demonstrates that single-epoch spectroscopy of RR~Lyrae stars is a viable method for probing the kinematics of the outer halo, and future surveys like Rubin/LSST and DESI-II have the potential to significantly advance this effort.

\end{abstract}

\keywords{catalogs --- Galaxy: halo --- methods: data  analysis --- stars: variables: RR Lyrae}

\newcommand{\kms}{km~s$^{-1}$}

\section{Introduction} \label{sec:intro}

The structure and kinematics of the Milky Way (MW) stellar halo provide a unique archaeological record of the Galaxy’s formation and evolutionary history. In the Lambda Cold Dark Matter ($\Lambda$CDM) paradigm of structure formation, the MW is thought to have assembled its dark matter halo over cosmic time through the accretion of hundreds of smaller dark matter halos, some of which hosted dwarf galaxies \citep{Bullock&Johnston2005}. The stellar debris from some of these accreted dwarfs is now observed in the MW's halo as substructures, both in spatial distribution \citep[e.g., tidal streams;][]{Belokurov2006} and in velocity–chemical-abundance space \citep[e.g., the Gaia–Sausage–Enceladus (GSE);][]{Vincenzo2019}. Recent studies have shown that the inner halo, within $r_{\rm GC} \lesssim 30$~kpc, is highly anisotropic \citep{Cunningham2019} and almost entirely comprised of merger remnants and substructures \citep{Naidu2020}. In the outer stellar halo ($r_{\rm GC} > 50$~kpc), simulations expect substructures to be even more dominant, as the longer dynamical timescales ($\gtrsim 1$~Gyr) allow accretion signatures to remain well preserved \citep[e.g.,][]{Sharpe2024}. This expectation is supported by observations of the outer stellar halo, including low spherically averaged radial dispersion \citep{Deason2012}, significant kinematic anisotropy \citep[][]{Han2024}, along with various identified substructures, such as the “Pisces Plume” \citep{Belokurov2019, Amarante2024} and the outer Virgo overdensity \citep{Sesar2017, Chandra2023}.

What adds further complexity to the study of the outer halo is the ongoing dynamical interaction between the MW and its most massive and luminous satellite, the Large Magellanic Cloud (LMC). The LMC is believed to have recently completed its first pericentric passage around the MW \citep{Kallivayalil2013, Sheng2024} and is currently located at a heliocentric distance of $\sim$50~kpc. Due to its substantial mass, the LMC exerts a significant gravitational influence on the MW, inducing a shift in the Galaxy’s center-of-mass (COM). While the disk and inner halo respond quickly to this shift and move as a whole due to their shorter dynamical timescales, the more distant outer halo stars—on multi-Gyr orbits—lag behind. This differential response results in a ``reflex motion'' signature observable in the outer halo kinematics, as demonstrated by simulations such as \citet{GC19} and \citet{Vasiliev2024}. These studies predict that outer halo stars will exhibit a dipole-like velocity pattern: stars in the northern Galactic hemisphere appear redshifted, while those in the south are blueshifted. \citet{GC19} also notably find that the amplitude of this dipole signal increases with Galactocentric distance, making the distant halo ($r_{\rm GC} \geq 100$~kpc) a particularly sensitive region for probing the LMC’s dynamical wake. 

Significant progress has been made by observers in detecting the dynamical influence of the LMC on outer halo stars in the MW. Several studies \citep{Peterson2021, Erkal2021, Yaaqib2024} have assembled samples of halo tracers—including K-giants, blue horizontal branch (BHB) stars, red giant branch (RGB) stars, RR Lyrae stars, and MW satellite galaxies—to measure the amplitude and direction (apex) of the predicted dipole velocity signal beyond 50~kpc. More recently, \citet{H3LMC} characterized the distance dependence of the dipole field using an all-sky sample of 850 RGB stars spanning 40 to 160~kpc. Similarly, \citet{DESILMC} analyzed 675 spectroscopically confirmed BHB stars between 50 and 120~kpc and reported dipole signatures consistent with those found in \citet{H3LMC}. While growing evidence from these studies supports that the kinematic imprints of the MW--LMC interaction become more prominent with increasing distance, a significantly larger number of kinematic measurements (at least $\sim$10 times more, as suggested by \citealt{H3LMC}) is still required to robustly characterize the dipole velocity field, particularly in the distant halo at $r_{\rm GC} \geq 100~\rm kpc$, where only a modest number ($\sim100$) of stars with published radial velocity measurements exist. In this work, we aim to expand the distant stellar halo sample by incorporating our new kinematic measurements of RR Lyrae stars.

RR Lyrae stars are pulsating variables with absolute magnitude $M_V \sim +0.6$~mag, low mass ($<1~M_{\odot}$), and short period ($<1$~d), ubiquitously found in old, metal-poor stellar populations \citep{Catelan2004}. They have long served as valuable tracers of the structure of the MW halo and disk \citep[e.g.,][]{Vivas2005,Belokurov2018,Li2022}. Compared to brighter giants, RR Lyrae stars offer unique advantages for probing the distant halo: (1) they follow a well-defined period--luminosity--metallicity relation in the optical bands \citep[e.g.,][]{17PS1, Garofalo2022}, enabling distance estimates with uncertainties as low as $\sim 5\%$ out to $\sim 300$~kpc—whereas BHB stars reach comparable accuracy only within $\sim 100$~kpc; (2) their distinct pulsation signatures allow for robust identification in time-domain photometric surveys, even with sparse sampling and irregular cadence. These strengths make RR Lyrae stars effective tracers of the distant halo, enabling their identification to large radii ($50 < r_{\rm GC} < 300$ kpc) with relatively well-measured distances and pulsational parameters \citep[e.g.,][]{17PS1, Stringer2021, NGVSRRL, Medina2024}. These stars thus become compelling targets for spectroscopic follow-up observations.

Radial velocity measurements of distant RR Lyrae stars are challenging for a combination of reasons: (1)~the stars are so faint that the S/N in the spectral continuum is relatively low even with long integrations on powerful telescopes; (2)~the combination of their warm effective temperature, low metallicity, and high surface gravity results in weak metal absorption lines; (3)~their short-term variability places practical limits on integration times and has other observational consequences. The stellar atmosphere undergoes continuous pulsation, causing the Doppler shifts of different absorption lines to vary differently throughout the pulsation cycle. Moreover, shock-induced emission features that emerge at specific phases further complicate the accurate determination of center-of-mass velocities \citep[e.g.,][]{Gillet2014}. Consequently, only a limited number of spectroscopic studies have targeted halo RR Lyrae stars for radial velocity measurements. These efforts typically rely either on densely sampled observations across the pulsation cycle \citep[e.g.,][]{2011for, Sesar2012, Liu2020, Medina25}, or on single-epoch observations timed to coincide with relatively quiescent pulsation phases, with center-of-mass velocities calibrated using empirical pulsation templates \citep[e.g.,][]{Medina23}. For RR Lyrae stars at $r_{\rm GC} > 100$~kpc, where comprehensive phase-resolved spectroscopy is even less feasible, the viable approach is to strategically schedule exposures on large telescopes, coupled with careful pulsation-phase corrections, to extract reliable velocity measurements.

In this work, we present our spectroscopic analysis of a sample of distant RR Lyrae candidates selected from photometric surveys, with the goal of extracting their kinematics and characterizing the velocity distribution in the context of the MW–LMC dipole perturbation model. In \S\,\ref{sec:data}, we describe the selection criteria and spectroscopic observations of our RR Lyrae candidate sample. In \S\,\ref{sec:analysis}, we describe the methodology used to derive the center-of-mass velocities of our targets. In \S\,\ref{sec:result}, we present the distribution of the Galactocentric velocities and perform model fitting to characterize the dipole velocity field. We discuss the implications of the derived velocity dispersion and dipole signal in the broader context of Galactic dynamics in \S\,\ref{sec:discussion}, and conclude with a summary of our findings in \S\,\ref{sec:summary}. The uncertainties of the radial velocities templates used in \S\,\ref{sec:analysis} and the spectroscopic identification of non-RR Lyrae targets are described in Appendices~\ref{sec:vel_model_err} and \ref{sec:bkgd_contam}, respectively.

\section{Data} \label{sec:data}
\subsection{Spectroscopic Target Selection} \label{subsec:TargetSelection}
A sample of 68 high-confidence distant RR Lyrae candidates in the heliocentric distance range $d_{\rm hel} \approx 85$--150 kpc was selected from the following sources: (1) 29 from NGVS \citep{Ferrarese2012,NGVSRRL}, (2) 20 from DES \citep{DES19}, and (3) 19 from PS1 \citep{17PS1}. Based on the published fitting scores in these surveys, we applied the following criteria to select RR Lyrae candidates for spectroscopic observation: (1) $s_{\text{fitting}} > 0.94$ for NGVS; (2) $s_1 > 0.9$, $s_2 > 0.9$, and $s_3 > 0.9$ for DES; and (3) $\text{score}_{\text{ab}} > 0.9$ for PS1. These fitting/identification scores are defined in \S\,3.1 of \cite{NGVSRRL}, \S\,4.3 of \cite{DES19}, and \S\,5.2 of \cite{17PS1}, respectively. It is important to emphasize that these RR Lyrae candidates were stringently selected with score thresholds higher than the suggested values in these papers, as we aim to get a less contaminated sample, which is located beyond the distances covered by most previous spectroscopic studies of the MW stellar halo. The identification of these RR Lyrae candidates and the estimation of their pulsational parameters (such as period, amplitude, and phase) are based on similar template fitting processes applied to multi-band light curve data. We predict the pulsation phases of these RR Lyrae candidates on our allocated observation nights to plan when to obtain their spectra (preferably within the phases between 0.35 and 0.8). We hope to capture these pulsating RR Lyrae stars during relatively quiescent, free-falling stages of their atmospheric pulsation cycle, as the emission components occurring in the Balmer line profiles at the expansion stage of the RR Lyrae atmosphere would affect the precision of our velocity measurement \citep{Gillet2014}. The ephemerides of observed RR Lyrae candidates are listed in Table \ref{tab:full}.  

The three time-domain photometric data sets---PS1, DES, and our own NGVS---from which we selected high-probability RR~Lyrae candidates for spectroscopy have limited/sparse cadence (30–-60 epochs) and moderate single-epoch photometric precision ($\sim0.05$--0.15~mag). Given this, it is expected that a fraction of even these high-probability RR~Lyrae candidates will turn out to be background contaminants such as quasars, active galactic nuclei, and blazars \citep[e.g., see \S\S\,3.3 and 3.6 of][]{NGVSRRL}; this is consistent with the results of our spectroscopic analysis presented in Appendix \ref{sec:bkgd_contam}. Even for objects that were correctly identified as RR~Lyrae, these same limitations of the photometric data sets can cause the pulsational parameters to be misestimated for a few reasons including: (a)~undetected quasi-periodic amplitude and period modulations associated with the so-called Blazhko effect \citep{1996A&A...312..111J} that could affect an estimated $\approx50$\% of the RR~Lyrae in our sample over the decade or so that has elapsed between the above photometric surveys and the Keck/ESI spectroscopy presented in this paper \citep{2017MNRAS.466.2602P, 2020MNRAS.494.1237S}; and (b)~misclassification of RRc (or even RRd) stars as RRab stars.

In fact, 4 of the 29 NGVS RR~Lyrae candidates that we selected for spectroscopy were initially classified as RRab. Subsequent improvements in our light curve fitting procedure led to these 4 stars being reclassified as RRc with period smaller than 0.3 day, and by then we had already carried out spectroscopic observations of these stars. Unlike the RRc/RRd vs.\ RRab misclassification cases mentioned in the last paragraph, these 4 stars have been analyzed correctly in the sense that we have used their best-fit RRc (not RRab) light curve information and velocity curve templates for our kinematical analysis.

\subsection{Spectroscopic Observations \& Data Reduction} \label{subsec:Obs}
Our spectroscopic data were obtained with the Keck~II 10-m telescope and the Echelle Spectrograph and Imager \citep[ESI;][]{2002PASP..114..851S} over a two-year period, from February 2020 to December 2022. In total, 13 nights were allocated by the Keck Time Allocation Committee (TAC); however, we were unable to observe on 2 nights due to bad weather and on another 1 night due to a ESI guider failure. We used the single-slit echelle mode of ESI, with $1 \times 1$ pixel binning and a slit width of $1\farcs0$ or $0\farcs75$ depending on seeing conditions, to target 68 RR Lyrae candidates. The weather- and instrument-related losses described above resulted in poor spectroscopic data quality for 3 of the targets. We drop these 3 targets from further consideration and focus on the 65 targets for which we were able to obtain spectra of adequate quality. The $1\farcs0$ or $0\farcs75$ configurations we used for this project result in a moderate spectral resolution level of $R$ $\sim 3900$ and $\sim 5200$, respectively. To determine the total exposure time for each RR Lyrae candidate, we set an $\text{S/N} = 5$ per pixel threshold at 6563~$\text{\AA}$, using each candidate's predicted photometric brightness and the respective ESI configuration for that night in the ESI exposure time calculator\footnote{\url{https://etc.ucolick.org/web_s2n/esi}}. This yielded total exposure times ranging from 1800~s (30 min) to 6000~s (1.67~hr), excluding readout time and overhead. We then divided the desired exposure time into individual exposures of 750, 900, and 1200~s to mitigate cosmic ray effects. The corresponding period coverage fractions of the observed RR Lyrae candidates range from \textbf{0.35} to \textbf{0.78}. We believe the absorption lines of interest should not significantly blur during the exposure time window we chose and affect our velocity measurements. The records of our ESI observation nights are shown in Table \ref{tab:esilogs}. 

Data reduction was performed using the ESIRedux tool in the \texttt{xidl} package developed by \citet[][see their \S\,2]{2003ApJS..147..227P}. This pipeline extracts the continuum on a order-by-order basis, and coadds the continuum over the multiple exposures of the same object. Each coadded order is then fluxed and stitched into a combined 1-D spectra. ESI has lowest throughput at order 6 and it results in the worst sky subtraction and S/N of all the orders. Since we are not using any line features over 9000 \text{\AA} for our kinematics analysis, we choose not to reduce order 6 for all our obtained spectra. Also, for 4 targeted RR Lyrae candidates that are unexpectedly faint and hard to extract any continuum in the blue orders, we choose to reduce only order 10 of their spectra for the quantitative brightness analysis described in the next section. In total, we have extracted coadded, and stitched 1-d spectra covering the wavelength range from 4000~\AA\ to 9500~\AA\ for 61 RR Lyrae candidates. The signal-to-noise ratio (S/N), measured within a $\pm75$~\AA\ window centered on 6563~\AA\ (with the absorption peak masked), ranges from 1.52 to 16.84, with the 33rd and 67th percentiles at 3.65 and 4.99, respectively. \textbf{The S/N measurements were performed using \texttt{specutils} \citep{specutils2019}, an Astropy-affiliated Python library \citep{astropy2018}.}

\begin{deluxetable*}{ccccc}
\tablenum{1}
\tablecaption{ESI Observation Records \label{tab:esilogs}}
\tablewidth{0pt}
\tablehead{
\colhead{Observation} & \colhead{Targeted RRab}  & \colhead{Seeing} & \colhead{Slit Width} &
\colhead{Full or Half} \\
\colhead{UT Date} & \colhead{Candidates}  & \colhead{(arcsec)} & \colhead{(arcsec)} &
\colhead{Night} 
}
\decimalcolnumbers
\startdata
2021Feb15 & 3 & 1.20 & 1.0 & Second Half\\
2021Feb16 & 5 & 1.00 & 1.0 & Second Half\\
2021Apr05 & 7 & 0.80 & 1.0 & Full\\
2021Apr06 & 6 & 0.70 & 0.75 & Full\\
2021Apr30 & 3 & 0.70 & 0.75 & First Half\\
2021May01 & 3 & 0.90 & 1.0 & First Half\\
2021Sep02 & 0 & - & - & Full\\
2021Oct31 & 7 & 0.75 & 0.75 & Full\\
2021Dec06 & 0 & - & - & Full\\
2022Feb04 & 8 & 0.90 & 1.0 & Full\\
2022Mar04 & 7 & 0.90 & 1.0 & Full\\
2022May02 & 0 & - & - & Full\\
2022Oct23 & 6 & 0.50 & 0.75 & Full\\
2022Nov25 & 8 & 0.60 & 0.75 & Full\\
2022Dec24 & 5 & 1.30 & 1.0 & Full\\
\enddata
\tablecomments{This table shows our observation record on ESI between 2021 February and 2022 December.}
\end{deluxetable*}

%\setlength{\tabcolsep}{3pt}
%\begin{longrotatetable}
\begin{deluxetable*}{ccccccccccccc}
\tablenum{2}
\tablecaption{RR Lyrae Kinematics and Uncertainties \label{tab:full}}
\tablewidth{0pt}
\tablehead{
\colhead{Object Name} & \colhead{RA} & \colhead{DEC} & \colhead{Period} & \colhead{gAmp} & \colhead{$R_{\rm GC}$} & \colhead{$T_{\rm obs}$} & \colhead{$\phi_{\rm obs}$} & \colhead{$t_{\rm exposure}$} & \colhead{S/N} &  \colhead{$v_{\text{sys}}$} & \colhead{$v_{\rm GSR}$}& \colhead{$\sigma_{\rm com}$}\\
\colhead{} & \colhead{(degree)} & \colhead{(degree)} & \colhead{(day)} & \colhead{(mag)} & \colhead{(kpc)} & \colhead{(MJD)} & \colhead{} & \colhead{(second)} & \colhead{} & \colhead{(\kms)} & \colhead{(\kms)} & \colhead{(\kms)} 
}

\startdata
DES-J030534.67-132110.10 & 46.39 & -13.35 & 0.54 & 1.21 & 143.49 & 59909.37 & 0.59 & 6000 & 2.78 & -14.93 & -66.24 & 10.96 \\
\enddata
\tablecomments{This table presents the results of the line-of-sight velocity extraction and center-of-mass velocity correction for the 55 RR Lyrae stars, as described in Section~\ref{sec:analysis}. Pulsational parameters and distances are adopted from their respective photometric survey papers. The complete table will be available in the online version in a machine-readable format. The machine-readable table will also include, the center-of-mass velocities measured from all four absorption features for each object, along with the breakdown of the systematic uncertainty $\sigma_{\rm com}$, into its components: $\sigma_{\rm cov}$, $\sigma_{\rm template}$, and $\sigma_{\rm correction}$}
\end{deluxetable*}
%\end{longrotatetable}

\begin{figure*}
    \centering
    \gridline{\fig{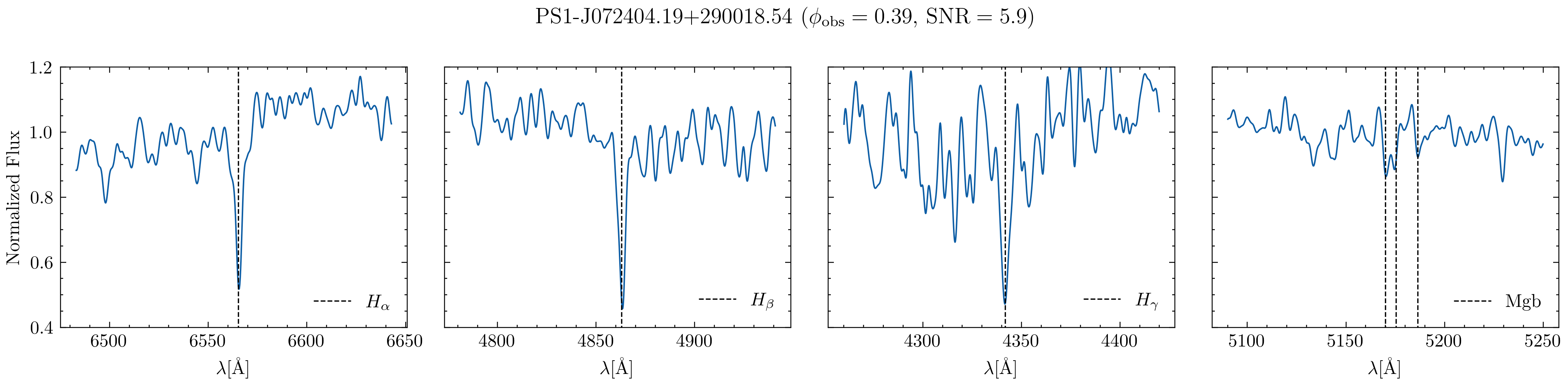}{0.95\textwidth}{(a)}
          }
\gridline{
          \fig{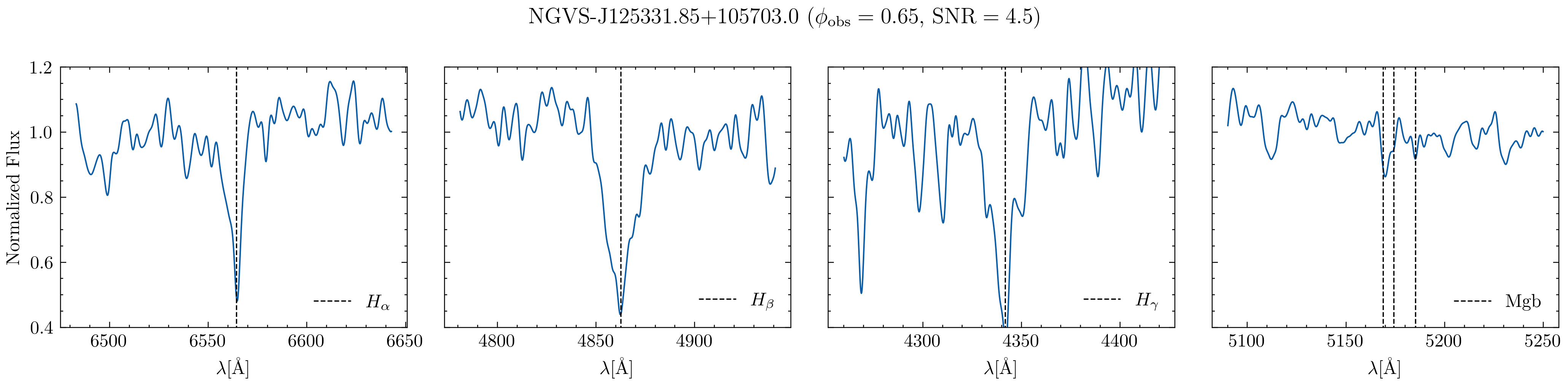}{0.95\textwidth}{(b)}
          }
\gridline{
          \fig{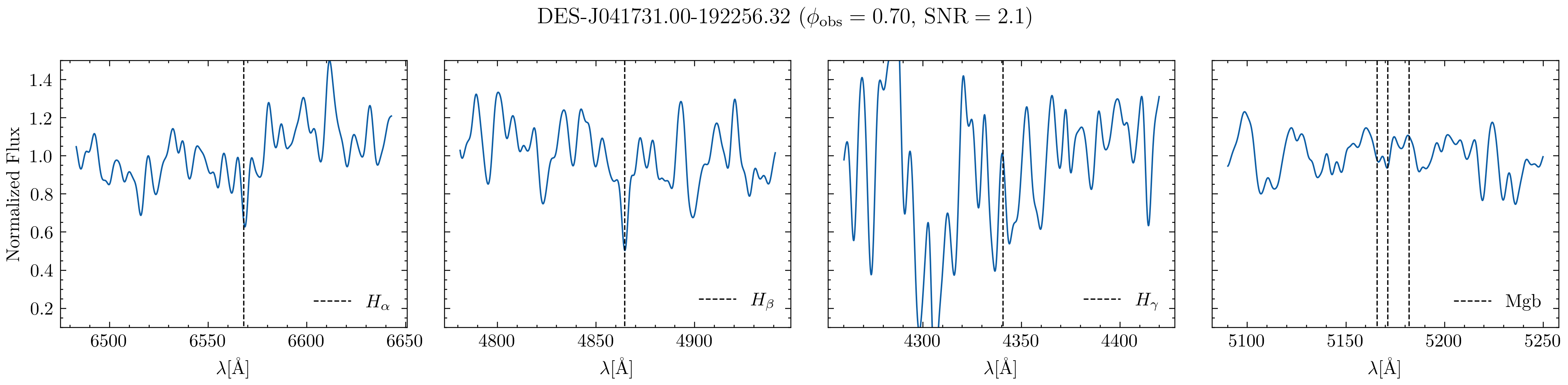}{0.95\textwidth}{(c)}
          }
%\gridline{
%      \fig{ngvs_2777_lines.png}{0.95\textwidth}{(d)}
%          }
    \caption{Sections of the spectra around the first three Balmer lines---H$\alpha$, H$\beta$, and H$\gamma$---and the Mg\,b triplet for three representative stars selected from the 55 RR Lyrae candidates used in our kinematic analysis. Our sample has 33rd and 67th percentile S/N values of 3.65 and 4.99 per \AA, respectively, and the three examples shown above represent typical S/N levels above the 67th percentile, between the 33rd and 67th percentiles, and below the 33rd percentile. The best-fit center wavelengths of these absorption features, as measured using the \texttt{pPXF} software, are marked by dashed vertical lines. For illustration purposes only, Gaussian smoothing with $\sigma = 1.5$~pixels has been applied, and each spectral segment has been normalized by dividing by the median flux in each spectral window.}
    \label{fig:StellarCases}
\end{figure*}

\begin{figure}
    \centering
    \includegraphics[width=0.95\linewidth]{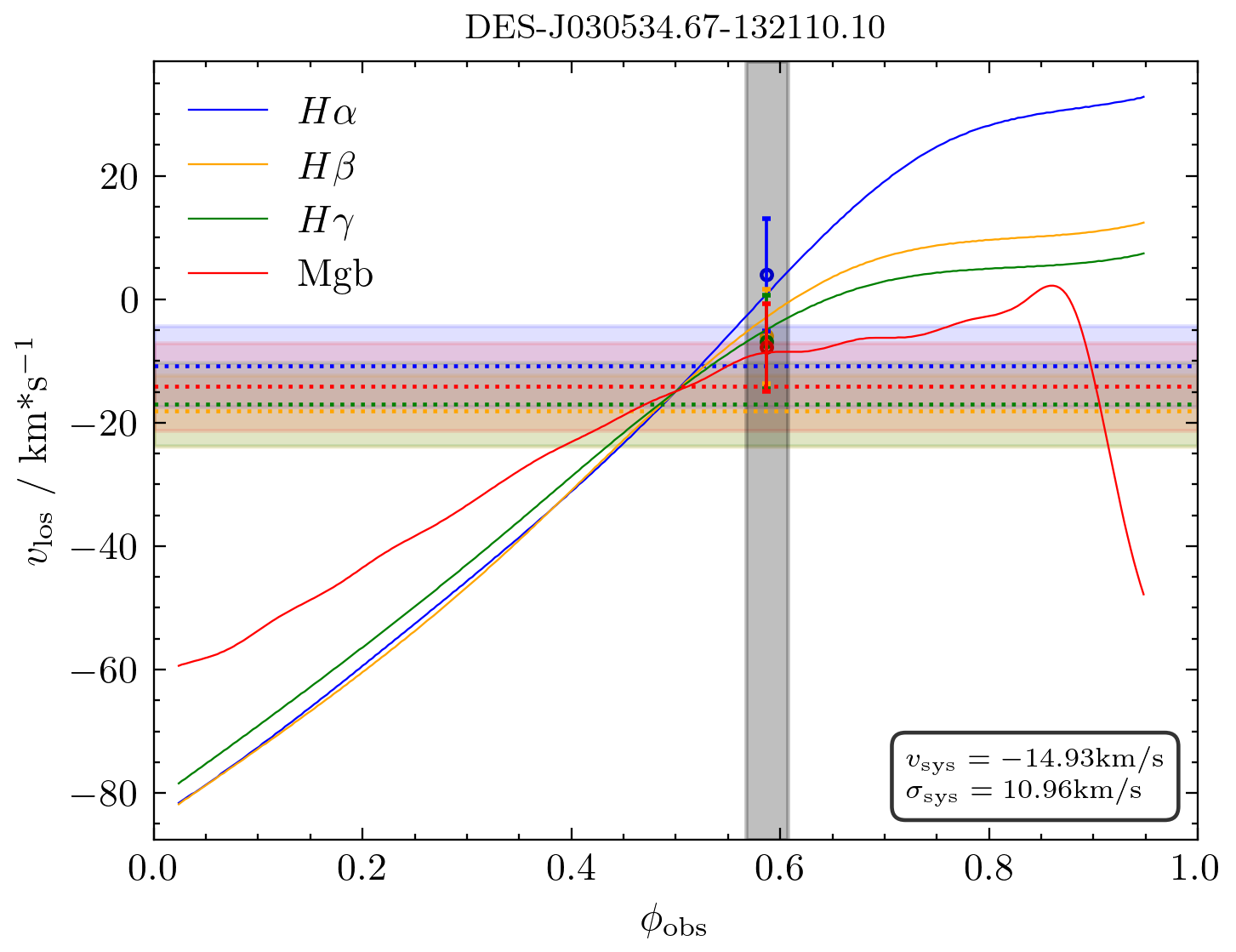}
    \caption{
    Example of the systemic velocity ($v_{\text{sys}}$) calculation for a single RR Lyrae star, based on Balmer and Mgb absorption lines. The radial velocity variation templates from \citet{Sesar2012} for H$\alpha$, H$\beta$, H$\gamma$, and Mgb are shown in blue, yellow, green, and red, respectively. The apparent line-of-sight velocities ($v_{\rm los,i}$) from each absorption feature are indicated by empty circles, with error bars representing the random uncertainties from the \texttt{pPXF} cross-correlation fitting. The corresponding center-of-mass velocities ($v_{{\rm com},i}$) derived from each feature are shown as horizontal dashed lines, with shaded bands indicating their individual uncertainties ($\sigma_{{\rm com},i}$). The gray vertical bar marks the phase coverage of the exposure sequence for this star. The final systemic velocity $v_{\text{sys}}$ and its total uncertainty $\sigma_{\text{sys}}$ are labeled at the lower right. This example illustrates that, by properly correcting for atmospheric pulsation and combining independent measurements of $v_{{\rm com},i}$, a reliable estimate of the systemic velocity can be achieved with reduced uncertainty.}
    \label{fig:vcal}
\end{figure}

\section{Analysis: RR Lyrae Velocity Extraction and Correction}\label{sec:analysis}

Of the 65 RR Lyrae candidates for which we obtained Keck/ESI spectra, 55 exhibit stellar absorption features such as H$\alpha$ (6563\,\AA), H$\beta$ (4861\,\AA), H$\gamma$ (4340\,\AA), and the Mg\,b triplet (5167~\text{\AA}, 5173~\text{\AA}, and 5183~\text{\AA}). These 4 absorption features were used in our cross-correlation analysis. It is noteworthy that we did not intend to get kinematical information from the Ca II triplet (at 8498~\text{\AA}, 8542~\text{\AA}, and 8662~\text{\AA}) throughout this study, mainly because ESI has lower efficiency at the red end from 8000 $\text{\AA}$ \citep[see Figure 13 in ][]{2002PASP..114..851S}. Combined with the fact that RR Lyrae stars are overall metal poor, the S/N around Ca II absorption region is too low to yield any reliable velocity extraction. Three examples of such spectra are shown in Figure~\ref{fig:StellarCases}, arranged in order of decreasing signal-to-noise ratio, from top to bottom: rows~(a)--(c). The remaining 10 targeted RR Lyrae candidates were identified as quasar or blazar contaminants based on our spectroscopic analysis; the details of that analysis are presented in Appendix \ref{sec:bkgd_contam}.

We utilized the \texttt{pPXF} cross-correlation routines \texttt{ppxf.ppxf} \citep[see][]{2012ppxf} to determine the apparent line-of-sight velocities of the 55 RR Lyrae candidates for subsequent spectral analysis. We selected 2 standard stars for the cross correlation analysis: HD 161817, a blue, metal-poor horizontal branch star for the cross correlation of Balmer lines; and BD +241676, a metal-rich F-type star with strong Mgb absorption feature specifically for the cross correlation of the metal triplet. We took ESI spectra for them with 2 second exposures during the twilight of our first Keck/ESI night. These bright stars have been extensively studied in the literature and have well-measured velocities by Gaia \citep{2020yCat.1350....0G}. Also, their absorption features (Balmer or Mgb) closely resemble those of RR Lyrae stars at quiescent phases. Around each absorption feature of interest (H$\alpha$, H$\beta$, H$\gamma$, and Mgb triplet), we extracted the nearby $\lambda_{\text{central}}\pm 75 \text{\AA}$ windows of the reduced 1-D spectra of the standard stars after shifting them to the rest frame, and cross-correlated them with the 1-D spectra of our science targets. The extracted line-of-sight velocities are shown in table \ref{tab:full}. 

When using RR Lyrae spectra for kinematic analysis, it is essential to subtract the atmospheric pulsation velocity $v_{\text{pul}}$ to derive the center-of-mass velocity, $v_{\text{com}}$. The amplitude and shape of the line-of-sight velocity curves vary depending on the spectral lines measured, primarily due to differences in the formation depths of these lines within the stellar atmosphere. Lines formed in the upper atmosphere, such as the Balmer lines, exhibit larger velocity variations compared to metallic lines formed deeper in the atmosphere \citep[e.g.,][]{2011for}. \cite{Sesar2012} has constructed template radial velocity curves for these absorption lines, from high precision spectral data throughout the pulsation cycle of RRab stars. We apply these empirical templates for the correction of the center-of-mass velocities of our observed RRab candidates for each line and each star. We converted the $g$-band photometric light curve amplitudes of our RRab candidates, which were reported by the photometric surveys, into $V$-band values using the relation $A_V = 0.9 A_g$. Then we rescaled the normalized velocity templates with physical velocity amplitudes, $A_{\text{rv,i}}$, using the empirical relationship between $A_V$ and $A_{\text{rv,i}}$, where the subscript i resembles one of the H$\alpha$, H$\beta$, H$\gamma$ and Mgb absorption features \citep[the converting relations between $A_{g},~A_V $ and $A_{\text{rv, i}}$could be seen \S5 and Eqs.~(2)--(5) of][]{Sesar2012}. For each given star and each absorption feature, we get the atmospheric pulsation velocity $v_{\text{pul,i}}$ at our observation epoch by plugging in the median phase value $\phi_{\text{obs}}$ of our exposure sequence on that RRab candidate into the rescaled velocity curve of feature i. The center-or-mass velocity, $v_{\text{com,i}}$, is obtained by subtracting $v_{\text{pul,i}}$ from the observed line-of-sight velocity $v_{\text{los,i}}$. For the 4 RRc stars we observed, a similar correction process was performed, with the RRc templates compiled by \cite{Braga21templates}. The parameters of the RRc templates and the conversion coefficients between $A_{\text{rv, RRc}}$ and $A_{\text{V, RRc}}$ could be found in their Table 10 and Table 17, respectively.

We obtained 4 independent center-of-mass velocities $v_{\rm com,i}$ from the 4 absorption features of a given star after the correction process described above. Figure~\ref{fig:vcal} illustrates how each line-of-sight velocity extracted using \texttt{pPXF} is corrected based on the pulsation model to obtain an independent measurement of $v_{\rm com, i}$. The final systematic center-of-mass radial velocity $v_{\rm sys}$ was calculated as the error-weighted average of the 4 ${v_{\rm com,i}}$. The error of each $v_{\rm com,i}$ is calculated as $\sigma_{\rm com,i} = \sqrt{\sigma_{\rm cov,i}^2+\sigma_{\rm template,i}^2 + \sigma_{\rm correction,i}^2}$, where the $\sigma_{\rm cov,i}$ is computed from the \texttt{pPXF} covariance matrix, the $\sigma_{\rm template,i}$ is reported by the respective paper of RR Lyrae radial velocity templates (see Appendix), and the heuristic correction term $\sigma_{\rm correction,i}$ we introduced is defined as: 
\begin{equation}
    \sigma_{\text{corr}} = a_{\rm i}\frac{A_{\text{rv},i}}{SNR} e^{b_{\rm i}|\phi-0.5|}
\end{equation}
where $a_{\rm i}$ and $b_{\rm i}$ are scaling coefficients to be determined for each absorption feature in H$\alpha$, H$\beta$, H$\gamma$ and Mgb. We believe this additional term is necessary to help us get a better evaluation of the systematic error because of the following three reasons:
 \begin{enumerate}
\item Following the approach of \citet{Sesar2012}, we perform cross-correlation ``blindly"—that is, without referencing simulated RR Lyrae line profiles at non-quiescent phases to account for the shock-front emission effects. These effects were incorporated in their analysis as a “template uncertainty” of the templates and propagated into the systemic velocity errors of our work, denoted as $\sigma_{\rm template,i}$ \citep[see Fig 6 of][]{Sesar2012}. However, the signal-to-noise ratios of our spectral sample are significantly lower than those in \citet{Sesar2012}, making our $v_{\rm los,i}$ measurements potentially more sensitive to line profile variations caused by shock-front effects in the RR Lyrae atmosphere. As a result, the $\sigma_{\rm template,i}$ values may be underestimated for our dataset, particularly for weak features such as the Mgb triplet—given the intrinsically low metallicity of RR Lyrae stars—and for the H$\gamma$ line, which lies in the noisy blue end of the ESI spectral range.

\item The pulsation parameters for our sample are derived from photometric surveys with limited cadence and moderate single-epoch precision. As a result, the calculated pulsation phase at the time of spectroscopic observation ($\phi_{\rm obs}$) cannot be assumed to be perfectly accurate. The uncertainty in $\phi_{\rm obs}$, though difficult to quantify directly, should be accounted for through an additional error term. \citet{Medina23} addressed this issue by allowing small shifts ($\pm 0.1$) in $\phi_{\rm obs}$ to minimize the scatter of ${v_{\rm com,i}}$ around $v_{\rm sys}$, while we choose to retain the $\phi_{\rm obs}$ values as provided by the photometric surveys and incorporate the resulting scatter into our error budget by increasing the uncertainties on the systemic velocity measurements.

\item Other factors-such as unaccounted-for cosmic rays, contamination from nearby sources, and variations in sky subtraction quality—can introduce additional uncertainties that vary across the wavelengths of the four absorption features. We account for these effects with an empirical error term that scales as a power law with the signal-to-noise ratio, specifically $\propto 1/\mathrm{SNR}$.
\end{enumerate}

%For the additional error term, we also made a natural assumption that the $\sigma_{\text{corr}}$ should increase proportionally with the RV pulsation curve amplitude $A_{\text{rv},i}$. We also assume that the further-off an RR Lyrae is from the quiescent phase (0.5), the exponentially more uncertain its velocity measurement is. We determined the $a_{\rm i}$ and $b_{\rm i}$ values with an iterative optimization process, by defining a measurement $\chi^2$ sum: \frac{1}{N}$\sum \frac{(v_{\rm com,i} - v_{\rm sys})^2}{\sigma_{\rm com,i}^2}$ over the $N = 55$ RR Lyrae candidates for each line feature i, and cost function $(\chi^2 -1)^2$ is minimized with \texttt{scipy.optimize.minimize} to get the best $a_{\rm i}$ and $b_{\rm i}$ values. In conclusion, our error correction operation increased the estimated error level of the final systematic velocity $v_\rm sys$ to better match the scatter level of the 4 $v_{\rm com,i}$, and it also made us add more weights to the $v_{\rm com}$ measured from H$\alpha$ and H$\beta$ when doing the weighted average, because they have smaller best-fit scaling coefficients $a$ and $b$. 

For the additional error term, we adopt the natural assumption that the correction uncertainty $\sigma_{\text{corr}}$ should scale proportionally with the amplitude of the radial velocity pulsation curve, $A_{\text{rv},i}$. We also assume that the uncertainty increases exponentially as the observation phase deviates from the quiescent phase ($e^{|\phi - 0.5|}$). We determine the values of scaling parameters $a_i$ and $b_i$ for each spectral line feature $i$ via an iterative optimization process. A normalized $\chi^2$ statistic summed over the $N = 55$ RR Lyrae candidates for each absorption feature $i$ is defined as:
\[
\chi_i^2 = \frac{1}{N} \sum_{k=1} \frac{(v_{{\rm com},i, k} - v_{\text{sys}, k})^2}{\sigma_{{\rm com},i, k}^2},
\]
and \textbf{the cost function $(\chi_i^2 - 1)^2$ was minimized using \texttt{scipy.optimize.minimize} \citep{SciPy2020} to obtain the best-fit values of $a_i$ and $b_i$.}

In conclusion, this error correction procedure increases the estimated uncertainty on the final systemic velocity $v_{\rm sys}$ to better reflect the observed scatter among the four $v_{{\rm com},i}$ measurements. It also leads to higher weighting of the H$\alpha$ and H$\beta$ measurements in the final weighted average, as they yield smaller best-fit scaling coefficients $a_i$ and $b_i$. The final $\sigma_{\rm sys}$ values—representing the uncertainties on the weighted-average center-of-mass velocities for our RR Lyrae kinematic sample—range from 6.44 to 35.63~\kms, with the 25th and 75th percentiles at 8.32 and 14.51~\kms, respectively. The velocity measurements and associated uncertainties for the 55 RR Lyrae stars included in our kinematic analysis are summarized in Table~\ref{tab:full}.

%% Putting eqnarrays or equations inside the mathletters environment groups
%% the enclosed equations by letter. For instance, the eqnarray below, instead
%% of being numbered, say, (4) and (5), would be numbered (4a) and (4b).
%% LaTeX the paper and look at the output to see the results.

\section{Results} \label{sec:result}
\subsection{Velocity Distribution of Distant Milky Way Halo RR Lyrae} \label{subsec: distribution}

\begin{figure}
    \centering
    \includegraphics[width=0.95\linewidth]{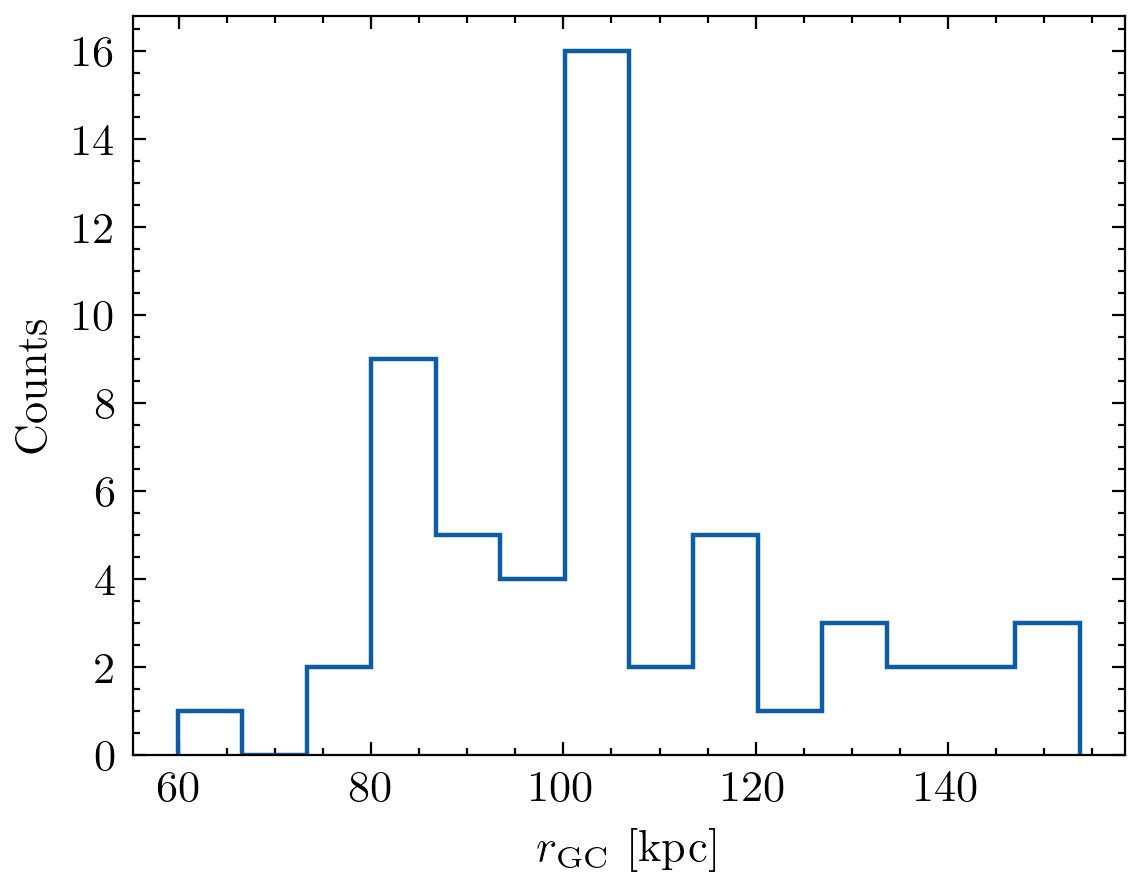}
    \caption{The Galactocentric distance distribution of the distant RR Lyrae kinematic sample after removing the contaminants described in Appendix \ref{sec:bkgd_contam}.}
    \label{fig:RGC_pdf}
\end{figure}

\begin{figure}
    \centering
    \includegraphics[width=1.0\linewidth]{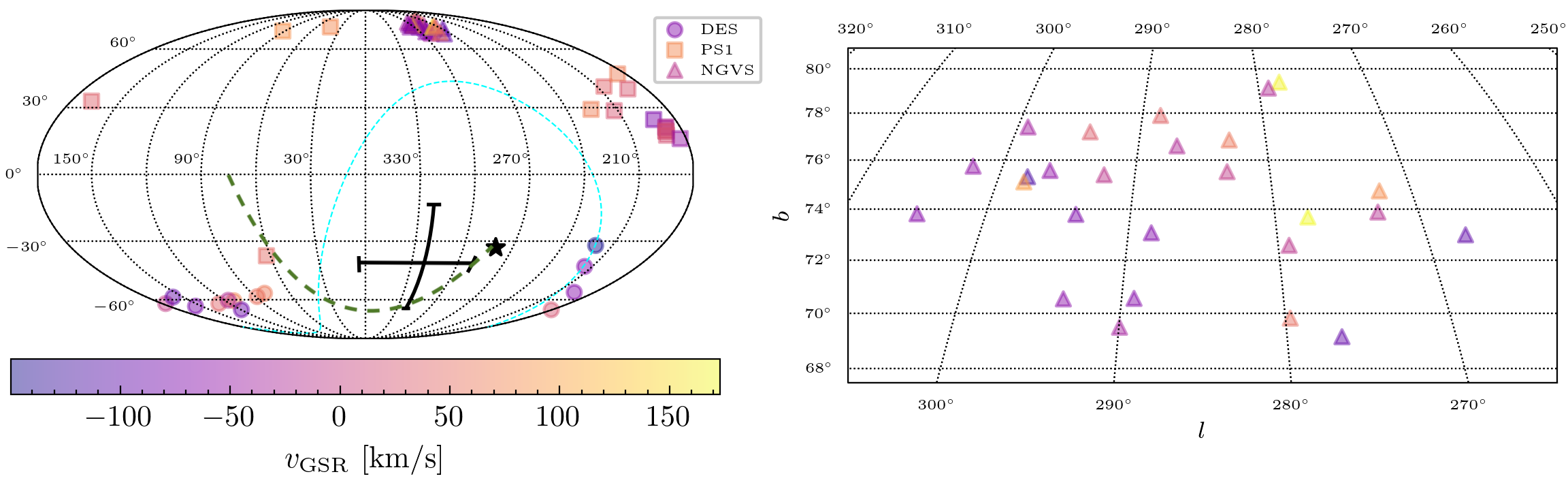}
    \caption{Individual $v_{\text{GSR}}$ measurements for stars in our dataset with their sky distribution. The current position of LMC is marked as a black star, and its simulated past trajectory (from the LMC3 model of \citet{GC19}) is shown as a green dashed line. The brown point with errorbars shows the apex position of our best-fit velocity dipole model from this sample (see \S \ref{subsec: Dipole}). The right panel is a zoom-in view of the sky distribution of our sample in the NGVS field.}
    \label{fig:mollweide}
\end{figure}

Our final sample consists of 55 RR Lyrae candidates with measured radial velocities and Galactocentric distances ($r_{\text{GC}}$) ranging from 60~kpc to 155~kpc. The distance distribution of this sample is shown in Figure~\ref{fig:RGC_pdf}. Figure~\ref{fig:mollweide} presents the spatial distribution of the sample in the Galactic Standard of Rest (GSR) frame, with each star color-coded by its Galactic radial velocity ($v_{\text{GSR}}$). We note that the filament-like structure along the Keck II declination limit ($\delta \approx -20^{\circ}$) in the southern hemisphere is an artifact because of the limited overlap between the visibility of the telescope and the DES footprint. No known halo substructures are present within the spatial and distance range covered by our sample.

\begin{figure}
    \centering
    \includegraphics[width=0.95\linewidth]{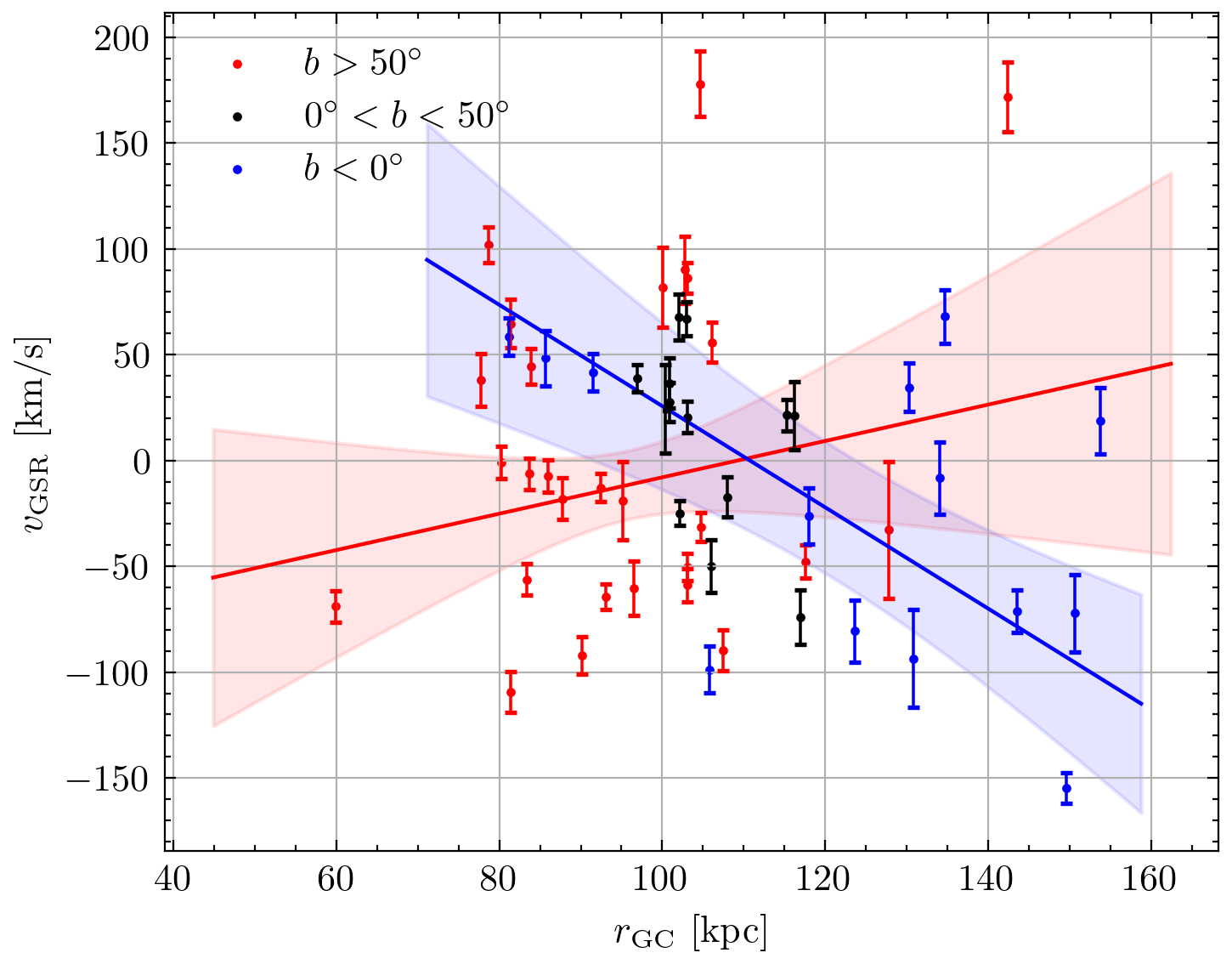}
    \caption{Scatter distribution of Galactocentric distances and $v_{\text{GSR}}$ of our sample. Linear regression results for the $b > 50^{\circ}$ and $b < 0^{\circ}$ subsamples are shown as red and blue lines, respectively. The corresponding 1$\sigma$ confidence intervals are indicated by the red and blue shaded regions.}
    \label{fig:scatter}
\end{figure}

\begin{figure}
    \centering
    \includegraphics[width=0.95\linewidth]{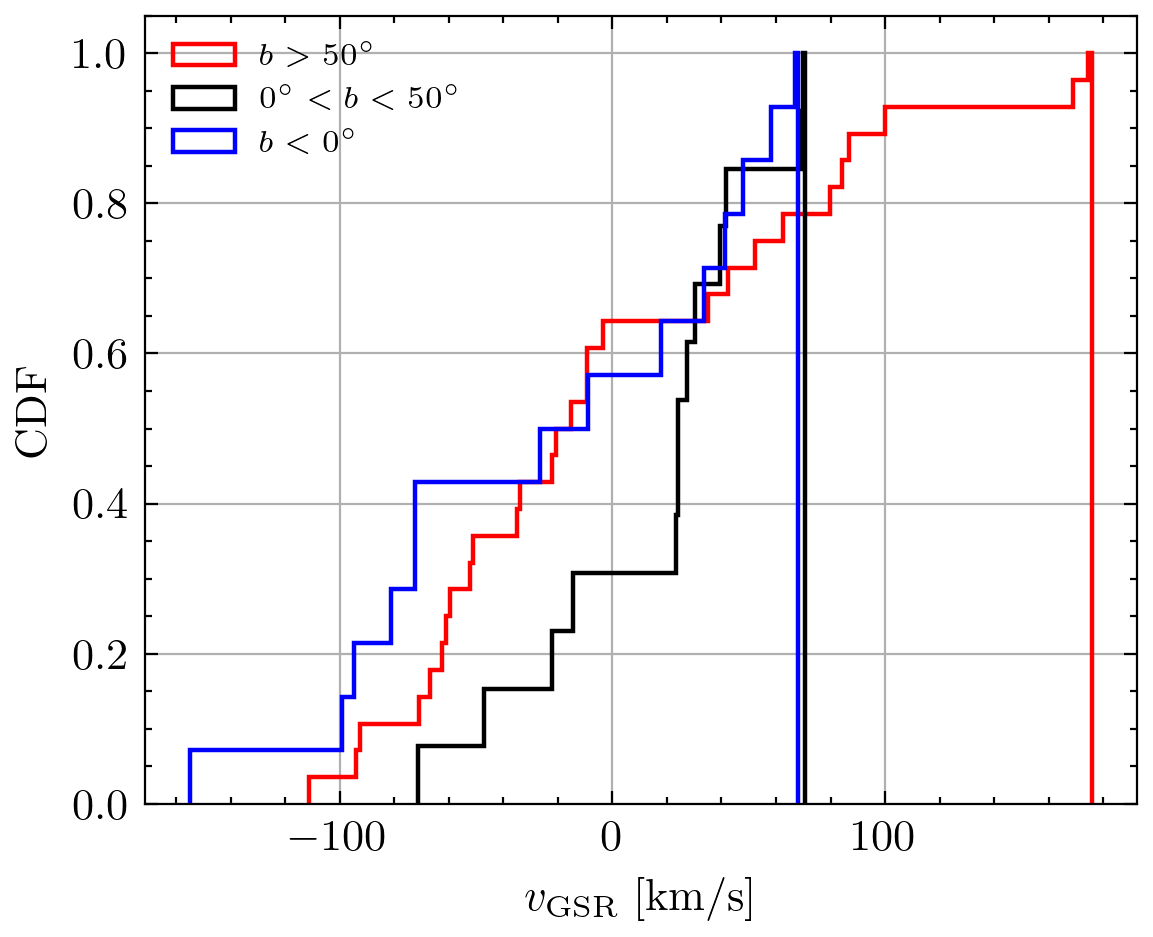}
    \caption{The cumulative distribution of the radial velocities in galactic standard frame for our RR Lyrae sample, grouped by their Galactic latitudes.}
    \label{fig:vel_cdf}
\end{figure}

We further investigate the structured distribution of $v_{\rm sys}$ by dividing our sample into three subgroups based on Galactic latitude:  
(1) 14 RR Lyrae candidates in the southern halo with $b < 0^{\circ}$;  
(2) 13 RR Lyrae candidates in the mid-latitude northern halo with $0^{\circ} < b < 50^{\circ}$; and  
(3) 28 RR Lyrae candidates in the high-latitude northern halo with $b > 50^{\circ}$.
Figure~\ref{fig:scatter} shows the distribution of $v_{\rm GSR}$ against their Galactocentric distances ($r_{\rm GC}$), and Figure~\ref{fig:vel_cdf} shows the cumulative distribution of $v_{\rm GSR}$ of the three subgroups. A weak dipole signature is visible in both the velocity map and the cumulative distribution, despite differences in distance coverage among the three subgroups. The high-latitude northern halo stars with $b > 50^{\circ}$ clearly exhibit a more red-shifted tail in their cumulative $v_{\rm GSR}$ distribution, while the southern halo subgroup has over 40\% of stars with $v_{\rm GSR} < -50$ \kms\ and a blue-shifted tail. We also performed linear regression on the $b > 50^{\circ}$ and $b < 0^{\circ}$ subsamples to study the correlation between $r_{\rm GC}$ and $v_{\rm GSR}$, obtaining best-fit slopes of $0.86 \pm 0.61$ $\rm km(s*kpc)^{-1}$ and $-2.31 \pm 0.55$ $\rm km(s*kpc)^{-1}$, respectively. This result is qualitatively consistent with the simulations of \citet{GC19}, where all of their 3 LMC models show that, the reflex motion response of the MW stellar halo becomes stronger with increasing distance \citep[see Figures~9–11 in][]{GC19}. However, the orientation of the reflex motion is not immediately evident in Figure~\ref{fig:mollweide}, mostly due to the limited sky coverage of our sample. We address this by presenting our best-fit multi-parameter dipole model in Section~\ref{subsec: Dipole} with our sample.

The velocity distribution also exhibits features that are not fully consistent with the simulation results of \citet{GC19}. First, while the southern stellar halo is systematically blue-shifted ($\overline{v_{\rm GSR}} = -25.5 \pm 9.7$~km~s$^{-1}$), the northern halo—particularly the $b > 50^{\circ}$ subgroup ($\overline{v_{\rm GSR}} = -9.2 \pm 3.3$~km~s$^{-1}$)—does not show the positive mean line-of-sight velocity predicted by \citet{GC19} at comparable Galactic latitudes (e.g. in their LMC3 model, $\overline{v_{\rm GSR}} > 35$~km~s$^{-1}$ at 100~kpc). This discrepancy is also evident in Figure~\ref{fig:vel_cdf}, where the $b > 50^{\circ}$ and $b < 0^{\circ}$ subgroups have nearly identical median $v_{\rm GSR}$ values at ${\rm CDF} = 0.5$. This may be attributed to unknown distant substructures in the Virgo direction, as our sampling at high Galactic latitudes is highly uneven: 26 out of the 28 high-latitude RR Lyrae candidates in the northern galactic hemisphere lie within the $\lesssim 110$~deg$^{2}$ footprint of NGVS.

Second, the overall $\overline{v_{\rm GSR}}$ of all 55 RR Lyrae candidates is $-12.2$~km~s$^{-1}$, indicating a net negative velocity offset from zero. While one model (LMC3) in \citet{GC19} also exhibits a similar negative all-sky average line-of-sight velocity, this signal is not fully explained by the dipole perturbation model alone.

Finally, the mid-latitude subgroup ($0^{\circ} < b < 50^{\circ}$) shows a significantly lower velocity scatter ($\sigma_{v_{\rm GSR}} = 40.7$~km~s$^{-1}$) compared to our all-sky sample ($\sigma_{v_{\rm GSR}} = 71.1$~km~s$^{-1}$), despite all models in \citet{GC19} predicting minimal perturbation from the MW-LMC interaction on the velocity dispersion level in this region (i.e., $\Delta \sigma_{\rm GSR} \approx 0$~km~s$^{-1}$; see their Equation~11). In other words, the mid-latitude subgroup is `colder' than other subgroups, which could not be explained by the dipole model either. We interpret this as possibly arising from small-number statistics, or more likely, the presence of unresolved halo substructures (see \S \ref{subsec:dispersion}).

%Figure~\ref{fig:mellow} immediately reveals that the outer halo exhibits a structured velocity distribution with significant variations across the sky. The Galactocentric radial velocity shows a net negative offset of approximately $-12$kms$^{-1}$. Additionally, there is a vague dipole signature in $v_{\text{GSR}}$, where the southern halo is slightly blueshifted and the northern halo is zero-centered.

%The reflex motion imparted by the LMC is evident in this $v_{\text{GSR}}$ map. Moreover, the orientation of the observed velocity signal is consistent with numerical simulations of the MW-LMC interaction. The bottom panel of Figure~\ref{fig:} presents the mean $v_{\text{GSR}}$ map from the MW+LMC simulations by \citet{GC19}, specifically their LMC3 model with an LMC virial mass of $1.8 \times 10^{11} M_\odot$. The observed velocity structure closely resembles the simulation results, particularly the tilt of the velocity dipole, though the observed dipole appears slightly stronger. Additionally, the simulation predicts a net negative sky-averaged $v_{\text{GSR}} \approx -10$\kms, suggesting that this signal may be induced by the LMC.

\begin{figure}
    \centering
    \includegraphics[width=0.98\linewidth]{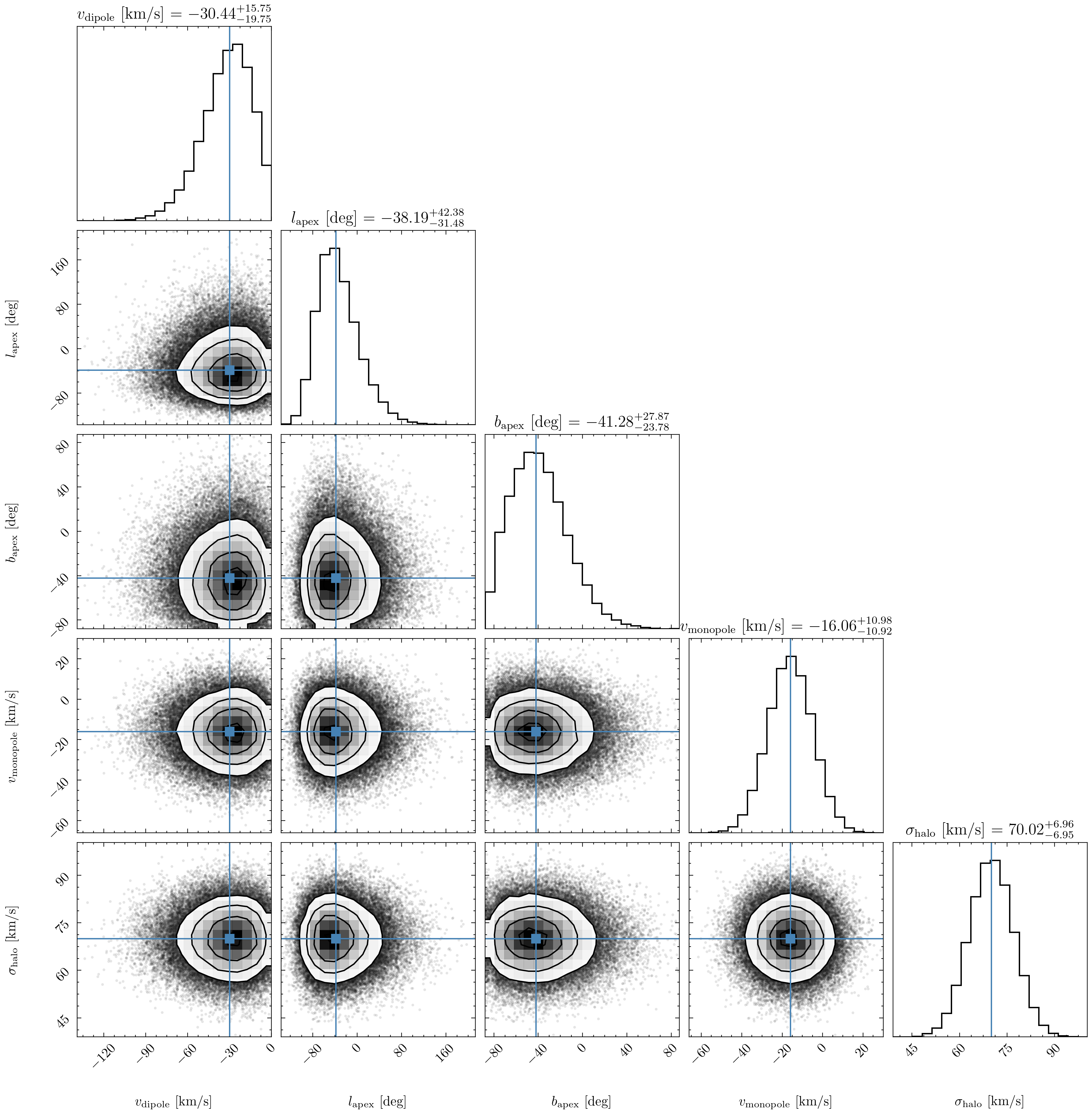}
    \caption{Corner plot showing the results of the Bayesian method of fitting a kinematical model consisting of the sum of a constant velocity offset (monopole term) and a velocity dipole.}
    \label{fig:corner}
\end{figure}

\subsection{Fitting an LMC Wake Model to Our Velocity Data} \label{subsec: Dipole}
We further model and analyze the radial velocity distribution of the RR Lyrae candidates in our sample as a function of Galactic coordinates, and examine the existence of the LMC wake effect.

First, we create a model under the assumption that their velocities could be characterized by a mean $\mu$ and intrinsic halo dispersion $\sigma_{\rm halo}$, before considering the dipole effect caused by the LMC. We account for individual uncertainties in the center-of-mass velocity measurements, $\sigma_{\rm com, k}$, for each observed RR Lyrae candidate denoted with k. The likelihood function could then be written in logarithmic format as:
\begin{align}
\label{eqa:MLE}
L &= \ln(P(\{v_{\rm GSR}, \sigma_{\rm com}\} | \mu, \sigma_{\rm halo})) \nonumber \\
  &\propto -\sum_{k=1}^{N} \left( \ln\left[2\pi(\sigma_{\rm halo}^2 + \sigma_{{\rm com},k}^2)\right] 
  + \frac{(v_{{\rm GSR}, k} - \mu)^2}{\sigma_{\rm halo}^2 + \sigma_{{\rm com},k}^2} \right)
\end{align}

We further parameterize the radial velocity field by modeling it as a dipole field. Instead of assuming $\mu$ being a constant, we re-express it as a function of its sky position:
\begin{equation}{\label{eqa:dipole}}
    \mu_k = v_{\text{monopole}} + v_{\text{dipole}}\cdot \cos(\theta_{k})
\end{equation}
where the $\theta_{k}$ is the angular distance between the given star at $(l_{k}, b_{k})$ and the unknown dipole apex $(l_{\text{apex}}, b_{\text{apex}})$. We fit the best estimated value of the five unknown parameters: $v_{\text{monopole}}$, $\sigma$, $l_{\text{apex}}$, $b_{\text{apex}}$, and the dipole strength $v_{\text{dipole}}$ by maximizing the likelihood function \ref{eqa:MLE} with the MCMC method. We employ the \texttt{emcee} sampler \citep{emcee} with 64 walkers and a total of 100,000 steps. We set the $v_{\rm dipole}$ to be negative in the MCMC process, and by geometric symmetry it would not bias the final result, since we allow an all-sky search for $l_{\rm apex}$ and $b_{\rm apex}$. The best fit result in shown in Figure \ref{fig:corner}. The MCMC fitting converged on a loosely constrained dipole model, yielding a reflex motion velocity of $v_{\rm dipole} = -30.44^{+15.75}_{-19.75}$~\kms\ and an apex direction of $(l_{\rm apex}, b_{\rm apex}) = (-38.19^{+42.38}_{-31.48}, -41.28^{+27.87}_{-23.78})$ degrees. Under this dipole model, we also obtained a monopole (bulk compression) velocity of $v_{\rm monopole} = -16.06 \pm 10.95$~\kms\ and a halo velocity dispersion of $\sigma_{\rm halo} = 70.02 \pm 6.95$~\kms. It needs to be mentioned that, we also tested a dipole-only model by excluding the monopole term. However, this led to degraded constraints on all parameters. In particular, the dipole amplitude became $v_{\rm dipole} = -23.57^{+29.02}_{-32.34}$~\kms, with ${\rm S/N} < 1$, effectively undermining the dipole interpretation. We therefore adopt the full dipole+monopole model and report the corresponding best-fit parameters. The direction of the dipole apex is marked on the left panel of Figure \ref{fig:mollweide}, and we could see its $1-\sigma$ confidence range marginally overlaps with the simulated trajectory of a $1.8\times10^{11}~M_{\odot}$ halo mass LMC from \cite{GC19}. We will further discuss our interpretation of these dipole parameters in the next section.

\section{Discussion} \label{sec:discussion}
\subsection{Radial Velocity Dispersion \& Effects from Potential Local Substructures}
\label{subsec:dispersion}

In this work, we have constructed a kinematic sample of distant halo RR Lyrae stars spanning Galactocentric distances from 60 to 160~kpc, with sparse sky coverage primarily at high Galactic latitudes ($|b| > 30^{\circ}$) in both the northern and southern hemispheres. We detect an overall velocity dispersion of $\sigma_{\rm halo} = 70.0 \pm 6.9$~\kms. Compared to previous simulations and measurements using other tracers, our result is broadly consistent but lies at the lower end of the expected range.

\citet{GC19} simulated halo radial velocity dispersion profiles based on the stellar density distributions of K giants \citep[][]{Xue2015} and RR Lyrae stars \citep[][]{Hernitschek2018}, yielding $\sigma_{\rm halo}$ in the range $80$--$90$~\kms\ at $r_{\rm GC} \sim 100$~kpc, which is more than 1$\sigma$ higher than our measurement. On the observational side, \citet{Deason2012} used CN stars and globular clusters to report a similarly low dispersion of $\sigma_{\rm halo} \sim 60$~\kms, although their result was limited by small-number statistics and distance uncertainties. More recently, \citet{H3LMC} compiled an all-sky sample of 850 RGB stars between 40 and 160~kpc and measured a dispersion of $\sigma_{\rm halo} = 75 \pm 5$~\kms\ at 100~kpc under a dipole model (see the lower panel of their Figure~10), which is consistent with the result of this study. In contrast, other recent studies focusing on the outer halo \textbf{within} 100~kpc have reported systematically higher values: \citet{DESILMC} found a radial dispersion of $90.5 \pm 2.5$~\kms\ for their sample of BHB  beyond 50~kpc, and \citet{Medina25} reported $\sigma_{\rm halo} = 137.3 \pm 2.0$~\kms\ for a metal-poor, non-Gaia-Sausage-Enceladus related RR Lyrae sample beyond 40~kpc. Such  differences in radial velocity dispersion between stellar samples within and beyond 100~kpc seems to support the scenario in which a substantial fraction of distant halo stars ($r_{\rm GC} \geq 100~\rm kpc$) are dynamically unrelaxed and belong to shell-like structures, sharing similar $v_{\rm GSR}$ as their progenitor galaxies \citep[e.g.,][]{Quinn1984}.

It should be noted that we do not attribute our low $\sigma_{\rm halo}$ measurement solely to the local substructures in the limited sky regions covered by this RR Lyrae sample. Firstly, while all LMC models in \citet{GC19} predict that MW--LMC interactions can induce localized “cold regions” with suppressed $\sigma_{v_{\rm GSR}}$, this effect is most pronounced in octants 5 and 6 of their Figure~11, near $l \sim 270^{\circ}$ and $b \sim 30^{\circ}$. The sky regions covered by our sample exhibit a relatively constant $\Delta \sigma_{v_{\rm GSR}} \sim 0$~\kms\ in the simulated models. In other words, the low $\sigma_{\rm halo}$ we report is not the result of localized LMC-induced kinematic perturbations. 

Also, while we do suspect that the mid-latitude subgroup ($0^{\circ} < b < 50^{\circ}$) may contain unresolved substructures \footnote{In the mid-latitude subgroup exist pairs of stars in close angular and spatial proximity with very similar $v_{\rm GSR}$, e.g., PS1-J072931.07+302920.80 and PS1-J072926.21+304524.19 with $(\Delta \theta, \Delta r_{\rm GC}, \Delta v_{\rm GSR}) = (0.23^{\circ}, 5.82~\text{kpc}, 7.65~\text{\kms})$;\\ PS1-J072404.19+290018.54 and PS1-J072313.10+290621.42 with $(2.01^{\circ}, 0.94~\text{kpc}, 0.29~\text{\kms})$}, removing the mid-latitude subgroup would not significantly increase the dispersion level. In fact, removing the entire subgroup raises the overall $\sigma_{\rm halo}$ only modestly, to $\sim 76$~\kms, which is still significantly lower than the $\sigma_{\rm halo}$ value reported by \citet{DESILMC} focusing on BHB stars mainly within 100 kpc. Therefore, we suspect that the low dispersion level is an intrinsic feature of the distant halo. We  retain these potential members of local substructures in our analysis, primarily due to the lack of reliable astrometric information at the faint magnitude limit ($g > 20.5$). Gaia DR3 does not provide proper motion measurements with $\text{SNR} > 1$ for any star in the mid-latitude subgroup, despite it being the brightest subgroup in our sample. We do encourage future follow-up astrometric observations to more decisively assess these stars' kinematic membership.
%In the mid-latitude subgroup exist pairs of stars in close angular and spatial proximity with very similar $v_{\rm GSR}$ (e.g., PS1-J072931.07+302920.80 and PS1-J072926.21+304524.19 with $(\Delta \theta, \Delta r_{\rm GC}, \Delta v_{\rm GSR}) = (0.23^{\circ}, 5.82~\text{kpc}, 7.65~\text{\kms})$;\\ PS1-J072404.19+290018.54 and PS1-J072313.10+290621.42 with $(2.01^{\circ}, 0.94~\text{kpc}, 0.29~\text{\kms})$). 

%We want to say that our $\sigma_{\rm halo}$ measurement is also unlikely to be the result of local fluctuations of radial velocity dispersion from the LMC perturbation. \citet{GC19} predicted that the MW-LMC interaction could cause a "local colder region" of radial dispersion at 100 kpc, but that 

\subsection{The Elusive LMC Wake}\label{subsec: LMCWakeEstimate}
The MCMC fitting result of the dipole model (Equation~\ref{eqa:dipole}) based on our sample, shown in Figure~\ref{fig:corner}, reveals a loosely-constrained, weak detection of the LMC wake signal. We compare our results with previous studies, including \citet{Peterson2021}, \citet{Yaaqib2024}, \citet{H3LMC}, and \citet{DESILMC}, which measured the same set of dipole parameters using outer halo tracers (e.g., RGB, K giants, and BHB stars) beyond 50~kpc. We interpret our fitted $v_{\rm dipole}$ and $v_{\rm monopole}$ as physically equivalent to their disc reflex motion velocities (or $v_{\rm travel}$) and compression velocities (or ``mean halo velocity"), respectively.

All these studies, including ours, yield estimates on $v_{\rm dipole}$ and $v_{\rm monopole}$ that are broadly consistent within the 1$\sigma$ confidence level. The results suggest that the MW disk is undergoing a reflex motion of $v_{\rm dipole} \approx 37$~\kms\ (weighted average of best-fit values from the literature), pointing toward the southern hemisphere ($b_{\rm apex} < 30^{\circ}$), while the outer stellar halo exhibits a net infall motion with $v_{\rm monopole} \approx -23$~\kms. However, the best-fit apex directions vary substantially, with $l_{\rm apex}$ ranging from $-75^{\circ}$ to $+60^{\circ}$ across different works. This spread highlights the inherent difficulty in precisely determining the LMC wake details through limited measurements of outer halo kinematics. 

Nonetheless, this study provides new insight into the kinematical state of the outer halo. Firstly, the outer stellar halo is clearly in a dipole disequilibrium due to the infall of the LMC. As shown in Figure~\ref{fig:scatter}, a systematic velocity asymmetry is observed between the northern and southern hemispheres, with blueshifted stars dominating in the south and redshifted stars in the north. Moreover, we could see the magnitude of the growth of $|\overline{v_{\rm GSR}}|$ increases more steeply in the southern hemisphere compared to the north, even without invoking a detailed dipole model. This behavior is predicted by all LMC interaction models in \citet{GC19} and \citet{Vasiliev2024}, and our results demonstrate that this effect persists out to beyond 100~kpc. Secondly, while we initially introduced the $v_{\text{monopole}}$ term in our dipole model to mathematically account for the observed amplitude difference between the northern and southern hemispheres, our fitting results favor the inclusion of this term and indicate that the compression component has a magnitude comparable to that of the dipole term $v_{\rm dipole}$, and is therefore non-negligible. The physical origin of such a large-scale compression across a wide distance range remains unclear, although a similar effect has been reported by nearly all recent observational studies of outer halo kinematics. \citet{Yaaqib2024} proposed that this compression could result from the gravitational pull of the LMC, which deepens the potential well of MW, and the most massive LMC model in \citet{GC19} does predict a net negative radial velocity field in the outer halo. However, given the limited and non-uniform sky coverage of our sample, it is also possible that this signal arises from unresolved substructures in phase space, potentially related to the anisotropy of the distant halo as shown in \citet{Amarante2024}. We anticipate that this effect will be better constrained through future observations and more comprehensive modeling.


\section{Summary} \label{sec:summary}
In this work, we conducted a spectroscopic study of RR Lyrae stars at Galactocentric distances of 60–160 kpc using Keck/ESI, to investigate the kinematics of the MW's outer stellar halo. We targeted 65 RR Lyrae candidates, selected from prior time-domain photometric surveys: NGVS, DES, and PS1. Spectroscopic analysis confirmed that 10 of these objects are non–RR Lyrae contaminants. For the remaining 55 candidates, we extracted line-of-sight velocities from their coadded spectra and corrected for center-of-mass velocities, achieving uncertainties ranging from 6.44 to 35.63 \kms.

We measured a velocity dispersion of $70.02 \pm 6.96$ \kms, indicative of a dynamically cold outer halo. The velocity field exhibits strong structure, with redshifted velocities in the northern hemisphere and blueshifted velocities in the south. Modeling the velocity field with a MW–LMC dipole perturbation model, we detect a weak dipole signal with a velocity amplitude of $-30.44^{+15.75}_{-19.75}$ \kms, pointing toward an apex direction of $(l_{\rm apex}, b_{\rm apex}) = (-38.19^{+42.38}_{-31.48}, -41.28^{+27.87}_{-23.78})$ degrees. This dipole may reflect the reflex motion of the MW disk due to the gravitational influence of the LMC. The fitting also supports the presence of a bulk compression velocity of $v_{\rm monopole} = -16.06 \pm 10.95$ \kms, the physical origin of which remains unclear.

Our dipole detection is not definitive, primarily due to the limited sky coverage and sample size. However, our results provide clear evidence that the LMC perturbation has placed the MW halo in a state of disequilibrium, with a dipole signature extending beyond 100 kpc. This study also demonstrates that single-epoch, moderate-resolution spectroscopy of distant RR Lyrae stars is a viable method for probing the outer halo. We expect that upcoming large-scale photometric surveys with sufficient depth \citep[e.g., the Vera C. Rubin Observatory;][]{LSST2009}, along with spectroscopic surveys such as the ongoing Dark Energy Spectroscopic Instrument (DESI) \citep{Schlafly2023} and the future DESI-II, will significantly expand the distant RR Lyrae kinematic sample, further our understanding of the MW outer halo, and characterize the imprints of the MW–LMC interaction.

\begin{acknowledgments}
YF and PG acknowledge support from National Science Foundation grant AST-2206328. We thank Meiyappa Chendrayan, Hannibal Dong, Calvin Sridhara, and Ethan Thomas for their help with Appendix\,\ref{sec:bkgd_contam}
of the paper; they participated in this research under the auspices of the Science Internship Program (SIP) at the University of California Santa Cruz.

\end{acknowledgments}


\facilities{Keck:II (ESI), CFHT (MegaCam)}

\software{\texttt{XIDL} \texttt{astropy}, \texttt{scipy}, \texttt{specutils}, \texttt{pPXF}, \texttt{emcee}}



\appendix

\section{RR Lyrae Velocity Template Error}\label{sec:vel_model_err}

The RRab velocity templates used in this work were constructed by \cite{Sesar2012}, which were defined to have a zero point at phase 0.5. This assumption is oversimplified, as the densely sampled time series radial velocity curves \citep[e.g.][]{2011for} have shown different atmospheric zero points. Therefore, the uncertainty $\sigma_{\rm template}$ for these templates was modeled as:
\begin{equation} \label{eqn: template}
    \sigma_{\rm template}^2 = \left( A_{\text{rv}} + \sigma_{\text{fit}} \right)^2 
\left[ \epsilon^2_{\text{template}}(\Phi_{\text{SDSS}}) + (0.1k)^2 \right]
\end{equation}
where the k is the slope of a template between phase 0.4 and 0.6. In the above equation, the term $\epsilon^2_{\text{template}}(\Phi_{\text{SDSS}})$ quantifies the root-mean-square (rms) scatter in atmospheric velocities around the normalized curve, measured from the original SDSS data used to construct the velocity templates. The term $\sigma_{\rm fit}$ accounts for the uncertainty introduced during the stacking and normalization of velocity curves from RR Lyrae stars with different $A_{\rm V}$ values in the template compilation process. The $(0.1k)^2$ term accounts for the uncertainty in the exact phase where the atmospheric velocity is zero. This term scales with both the slope of the template near the zero-velocity point and the velocity pulsation amplitude $A_{\rm rv}$. These parameters are reported in \S\,6 of \citet{Sesar2012}. Following the calculation of equation \ref{eqn: template}, the over-simplified zero-point assumption introduces systematic errors of approximately $15$\kms\ to Balmer line radial velocities for a typical RRab star with $A_V \sim 1$~mag, and around 7~\kms\ for Mgb velocities of the same star. Nonetheless, the zero-point error can be significantly reduced by simultaneously using the velocity measurements from multiple absorption features. For our sample, the maximum synthesized template uncertainty from all 4 features is below 6~\kms.

For the RRc templates from \citet{Braga21templates}, the zero point is not fixed at phase 0.5. Therefore the template uncertainty can be directly calculated as $\sigma_{\rm template} = A_{\rm rv} \sigma_{\rm fit}$ by rescaling the dimensionless uncertainties ($\sigma_{\rm fit}$) provided in the last column of their Table~10.

\section{Background Contaminants}\label{sec:bkgd_contam}
\begin{figure*}
    \centering
    \gridline{\fig{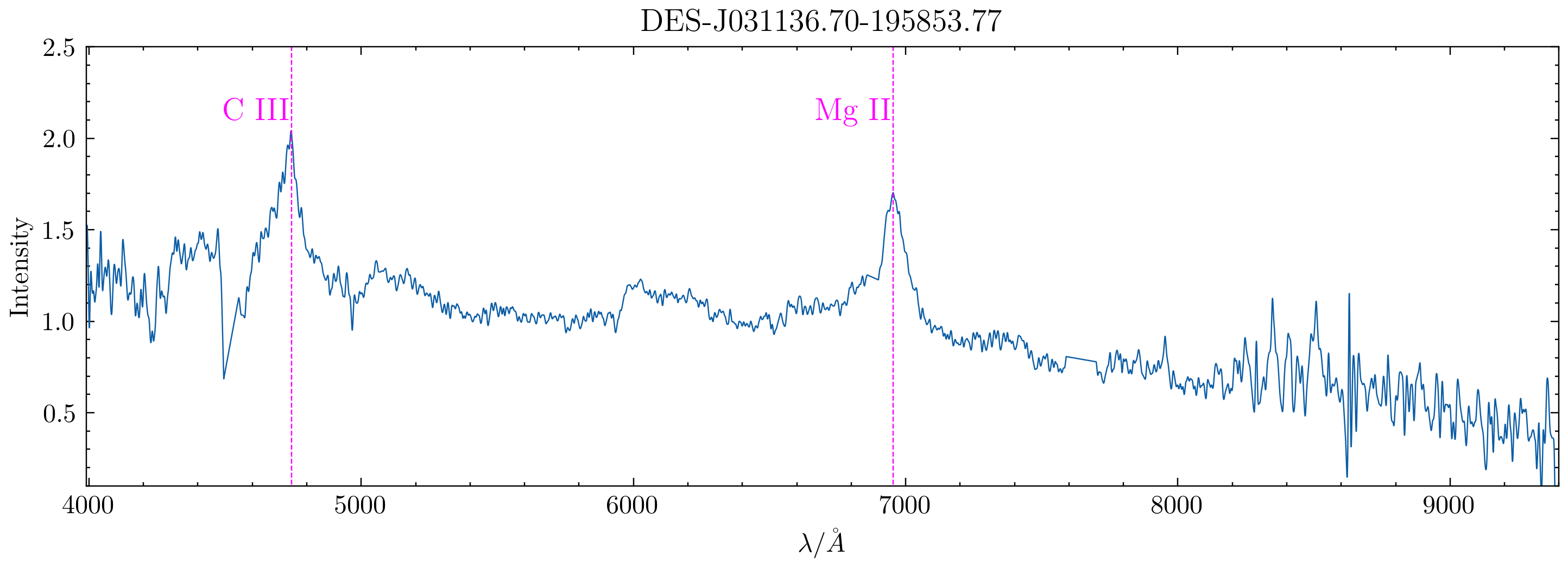}{0.95\textwidth}{(a)}
          }
\gridline{
          \fig{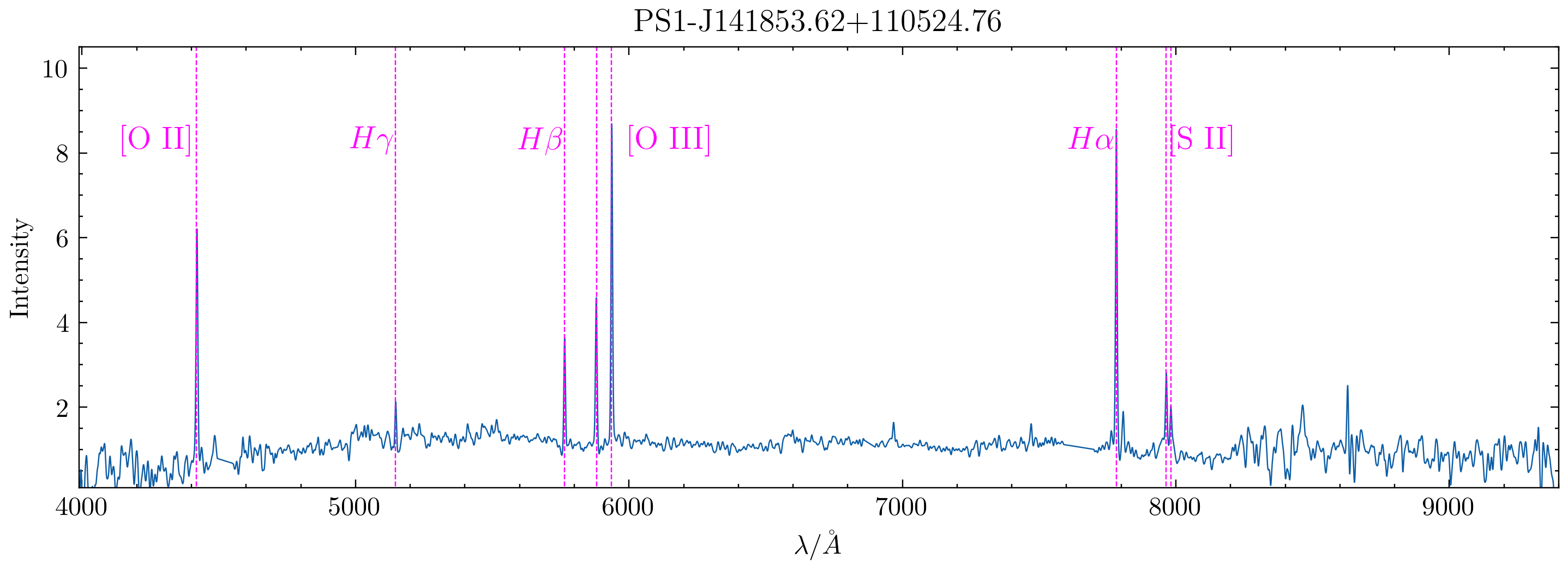}{0.95\textwidth}{(b)}
          }
\gridline{
          \fig{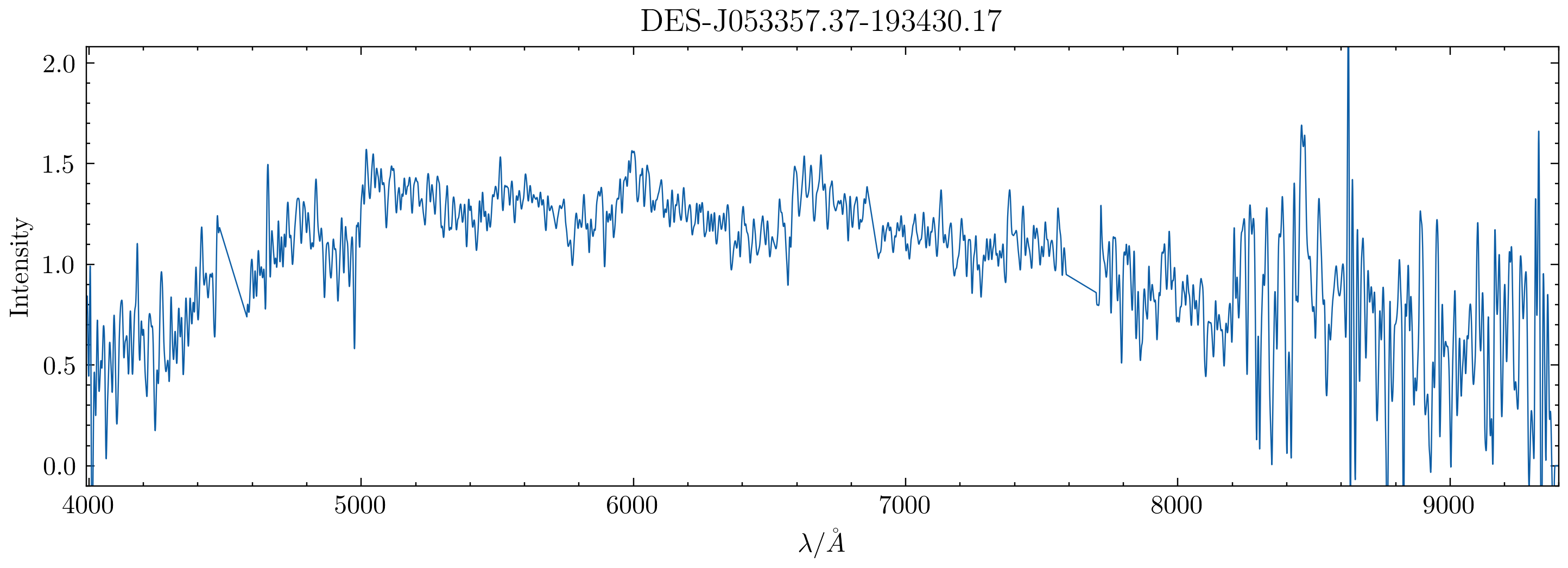}{0.95\textwidth}{(c)}
          }
    \caption{Showcases of full spectra of three RRab candidates that exhibit non-stellar features and were excluded from our kinematic analysis. (a) A quasar with broad emission lines from C~III and Mg~II, with a best-fit redshift of $z = 1.484$; (b) an object showing strong narrow emission lines at [O~II], H$\alpha$, H$\beta$, [O~III], and [S~II], likely an AGN or star-forming galaxy, with a best-fit redshift of $z = 0.186$; (c) an object with no discernible stellar features, possibly a blazar misidentified as an RRab. The visible steps in the continuum are a result of imperfect order stitching. The wavelength regions affected by the atmospheric O$_2$ absorption bands, as well as an instrumental artifact due to defective CCD columns in ESI (4490--4520~\AA), were masked and the flux was linearly interpolated over the resulting gap. Spectra images of other objects in Table \ref{tab:excluded} are available in the online version.}
    \label{fig:EmissionCases}
\end{figure*}

As described in \S \ref{sec:analysis}, we find that 10 targeted RR Lyrae candidates are unlikely to be genuine RR Lyrae stars based on their spectral properties. Of these, six objects have good-quality, fully reduced spectra that exhibit clear non-stellar features, such as emission lines (QSO contaminants) or featureless spectra (potential blazars). Three typical examples are shown in Figure~\ref{fig:EmissionCases}.Additionally, four candidates appeared abnormally faint, with no visible continuum after their initial $\sim$20-minute exposures. We also consider these objects to be potential long-term variables or transient contaminants, such as QSOs or blazars, that became significantly fainter after the initial DES or PS1 photometric observations. Their exposure sequences were aborted after the first 20 minutes. Below, we present a quantitative analysis supporting the conclusion that these objects are unlikely to be genuine RR Lyrae stars with well estimated pulsational parameters, by estimating their apparent brightness from their continuum levels in ESI order 10.

For each observed candidate, we extracted the 10th spectral order (covering 5620--6570~\AA) from their first exposure. This order was chosen because some objects were too faint to show strong continuum in the red, low-sensitivity end of ESI; for high redshift objects with non-stellar SEDs, their low continuum levels at the blue end also prevented full-order extraction. We measured the spectrum intensity at 5900~\AA, denoted as $F_{5900}$. To estimate the intensity at the central wavelengths of the SDSS $g$ band (4770~\AA) and $i$ band (7630~\AA), we extrapolated $F_{5900}$ using the continuum profile of the RR Lyrae candidate PS1J070856.46-031531.97, a $g \sim 13$ bright RRab we observed at phase $\phi = 0.50$ during one twilight of our ESI nights, obtaining estimated flux values of $F_{4770}$ and $F_{7630}$. We then derived spectroscopic magnitudes $g_{\text{spec}}$ and $i_{\text{spec}}$ for each candidate using the relations $g_{\text{spec}} = -2.5\log{F_{4770}} + g_0$ and $i_{\text{spec}} = -2.5\log{F_{7630}} + i_0$, where the constants $g_0$ and $i_0$ were determined from the predicted photometric $g$ and $i$ magnitudes of PS1J070856.46-031531.97 at phase 0.5, based on the pulsational parameters reported in \citet{17PS1}.

\begin{figure}
    \centering
    \includegraphics[width=0.95\linewidth]{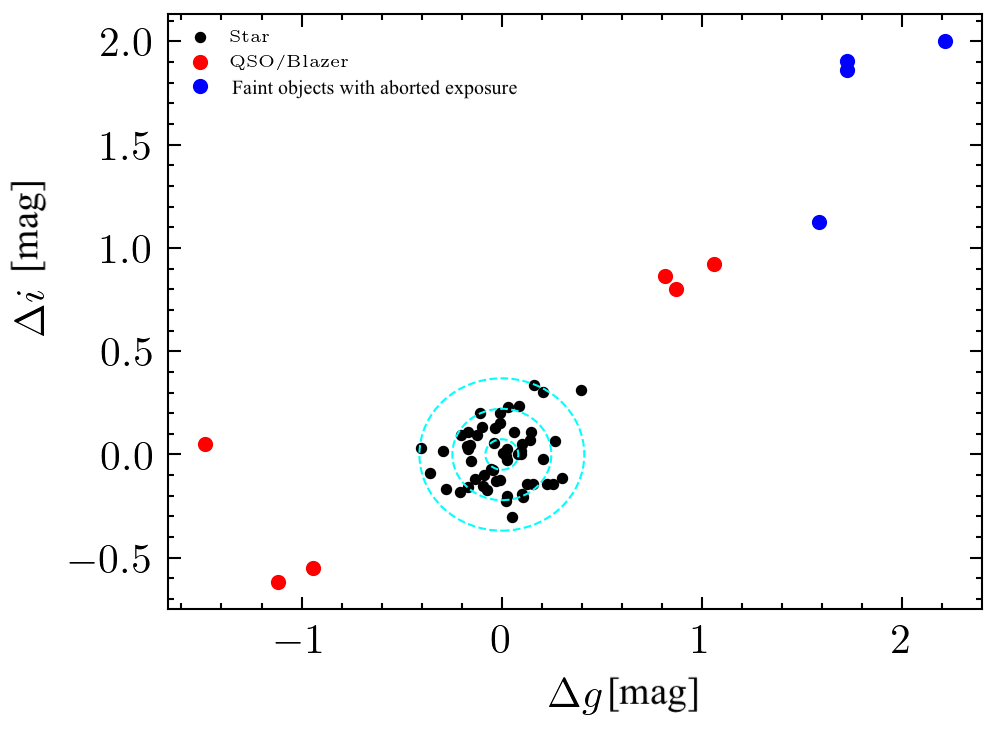}
    \caption{The $\Delta g$ vs. $\Delta i$ distribution of all 65 observed RR Lyrae candidates. Three subgroups are highlighted: (a) stellar objects exhibiting clear Balmer absorption features (black circles), (b) non-RR Lyrae objects showing strong emission lines in their full spectra (red circles), and (c) abnormally faint objects whose exposure sequences were aborted after the initial 20-minute exposure (blue circles). Cyan ellipses denote the 0.5$\sigma$, 1$\sigma$, and 2$\sigma$ scatter contours of the stellar-object distribution in both $\Delta g$ and $\Delta i$ directions, centered on the origin. All targets with non-stellar emission features and the faint aborted objects are clearly offset from the main stellar locus.
}
    \label{fig:dgdi}
\end{figure}

The estimated spectroscopic magnitudes were then compared to the predicted model-based magnitudes $g_{\text{phot}}$ and $i_{\text{phot}}$, calculated from pulsational parameters reported in the three literature papers. The distribution of the deviations, $\Delta g = g_{\text{spec}} - g_{\text{phot}}$ versus $\Delta i = i_{\text{spec}} - i_{\text{phot}}$, is shown in Figure~\ref{fig:dgdi}. As illustrated, targets with stellar absorption features loosely cluster near the origin, while those showing QSO or blazar-like features are more widely scattered. The faint targets, whose exposure sequences were aborted after the initial 20 minutes, are indeed abnormally faint and located in the upper right corner of the plot, suggesting that their reported pulsational parameters—or even their classifications as RRab stars — may be unreliable. 

%%%%%It should be noted that the parameters $\Delta g$ and $\Delta i$ we introduced above are not accurate tracers for the goodness of RRab classification and characterization of pulsation features, due to the following reasons
We further discuss the potential error sources of this test, and try to quantitatively explain the scattering of stellar targets around the origin point. On the one hand, our estimates of $g_{\text{spec}}$ and $i_{\text{spec}}$ are not fully calibrated, as effects such as atmospheric transparency, instrument response variations, nearby light contamination, and cosmic rays were not corrected in this simplified analysis. Also, the $T_{\text{eff}}$ and shape of a stellar SED varies throughout the pulsation cycle of an RRab \citep[e.g. see Fig 6 in][]{Medina25}, introducing further inaccuracy to the flux extrapolation step. On the other hand, the $g_{\text{phot}}$ and $i_{\text{phot}}$ reported from photometric surveys could also be biased \citep[e.g.][which reported remarkably different period estimations for $\sim$ 10\% of overlapping PS1 RR Lyrae in K2 survey]{K2PS1check}, because of their limited/sparse cadence and moderate single-epoch precision, as discussed in \S \ref{subsec:TargetSelection}. Therefore, within the loose cluster around the origin, we cannot further assess the classification reliability of the RR Lyrae candidates based on their exact $\Delta g$ and $\Delta i$ values, nor can we rule out the possibility that non-RR Lyrae stellar contaminants (e.g., eclipsing binaries) coincidentally fall within the cluster. 

Nevertheless, all the error sources discussed above propagated to a scatter of only 0.14 mag among the 55 stellar objects around the origin point, and we believe that the 10 objects exhibiting large ($\geq 1$ mag) deviations in the $\Delta g$ and $\Delta i$ distribution are highly likely to be non-RR Lyrae contaminants. This quantitative test shows corroborating results to support our rationale for excluding these problematic objects from our subsequent kinematic analysis of the stellar halo.

We summarize the list of excluded objects in Table \ref{tab:excluded}, along with the reasons for their removal, and we encourage future follow-up spectroscopic or photometric observations to confirm their classifications and pulsational parameters.

\begin{deluxetable*}{cccccccc}
\tablenum{3}
\tablecaption{Excluded Non RR Lyrae Objects \label{tab:excluded}}
\tablewidth{0pt}
\tablehead{
\colhead{Object Name} & \colhead{RA} &\colhead{DEC} &\colhead{Period} &\colhead{gAmp} &\colhead{Classification Score} &\colhead{Observed Spectroscopic Feature} & \colhead{Estimated Redshift}\\
\colhead{} & \colhead{(degree)} & \colhead{(degree)} & \colhead{(day)} &\colhead{(mag)} & \colhead{(Classifier 1/2/3)} & \colhead{} & \colhead{}
}
\startdata
DES-J031136.70-195853.77 & 47.90 & -19.98 & 0.7359 & 0.71 & 1.0/0.94/0.93 & C III and Mg II broad emission lines & 1.48\\
DES-J022346.15-012843.94 & 35.94 & -1.48 & 0.6489 & 0.69 & 0.97/0.94/0.91 & C III and Mg II broad emission lines & 1.31\\
DES-J053357.37-193430.17 & 83.49 & -19.58 & 0.5947 & 1.23 & 1.00/1.00/1.00 & Featureless (Blazar?) & -\\
DES-J054301.12-215937.63 & 85.75 & -21.99 & 0.5806 & 1.02 & 0.90/0.94/0.96 & Unreasonably Faint & -\\
DES-J004324.70-191552.26 & 10.85 & -19.26 & 0.6647 & 0.39 & 0.96/0.90/0.91 & Unreasonably Faint & -\\
DES-J011827.43+023048.79 & 19.61 & 2.51 & 0.6064 & 0.73 & 0.90/0.94/0.95 & Unreasonably Faint & -\\
DES-J052950.94-230708.78 & 82.46 & -23.12 & 0.5299 & 1.20 & 0.90/0.90/0.98 & Unreasonably Faint & -\\
PS1-J141853.62+110524.76 & 214.72 & 11.09 & 0.5718 & 1.15 & 0.90 & [OII], [OIII], [NII], [SII] emissions& 0.19\\
PS1-J184930.97+430523.93 & 282.38 & 43.09 & 0.5481 & 0.60 & 0.90 & C III and Mg II broad emission lines & 1.21\\
PS1-J173233.74+340209.38 & 263.14 & 34.04 & 0.6383 & 0.99 & 0.90 & Featureless (Blazar?) & -\\
\enddata
\tablecomments{This table shows the objects which were photometrically observed and classified as RRab candidates but were excluded from our kinematics analysis based on the spectroscopic analysis described in Appendix \ref{sec:bkgd_contam}. The best-fit periods, $g$ band amplitudes and scores of RRab classification are cited from their respective photometric survey papers. This table will be available in the online version in a machine-readable format.}
\end{deluxetable*}


\bibliography{ESI_Spec}{}
\bibliographystyle{aasjournal}



\end{document}